\documentclass[preprint,aps,prc,superscriptaddress,showpacs,nofootinbib,floatfix]{revtex4}
\usepackage{graphicx} \usepackage{dcolumn} \usepackage{bm}

\newcommand{\beq}{\begin{equation}}\newcommand{\eeq}[1]{\label{#1}
\end{equation}}\newcommand{\beqar}{\begin{eqnarray}}\newcommand{\eeqar}[1]
{\label{#1}
\end{eqnarray}}\newcommand{\bmath}{\begin{displaymath}}\newcommand{\emath}{\end{displaymath}}\newcommand{\bitem}{\begin{itemize}}\newcommand{\eitem}{\end{itemize}}

\begin{document}

\title{\Large \bf Open Charm Production in $p + p$ and Pb + Pb collisions 
at the CERN Large Hadron Collider.}

\newcommand{\mcgill}{Physics Department, McGill University, Montreal, Canada, H3A 2T8}

\newcommand{\columbia}{Physics Department,
Columbia University, New York, N.Y. 10027}

\newcommand{\ifin}{National Institute for Physics and Nuclear Engineering-Horia~Hulubei, 
R-077125, Bucharest, Romania}

\affiliation{\mcgill}
\affiliation{\columbia}

\author{~V.~Topor~Pop} \affiliation{\mcgill}
\author{~M.~Gyulassy} \affiliation{\columbia}
\author{~J.~Barrette} \affiliation{\mcgill}
\author{~C.~Gale}  \affiliation{\mcgill}
\author{~M.~Petrovici} \affiliation{\ifin}

\date{December 6, 2013}

\begin{abstract}

Effects of strong longitudinal colour electric fields (SCF), shadowing, and quenching   
on the open prompt charm mesons (D$^0$, D$^+$, D$^{*+}$, D${_s}{^+}$) 
production in central  Pb + Pb collisions at $\sqrt{s_{\rm NN}}$ = 2.76 TeV  are investigated 
within the framework of the {\small HIJING/B\=B v2.0} model.
We compute the nuclear modification factor $R_{\rm PbPb}^{\rm D}$, 
and show that the above nuclear effects constitute important 
dynamical mechanisms in the description of experimental data. 
The strength of colour fields (as characterized by the string tension $\kappa$), 
partonic energy loss and jet quenching process lead to a suppression factor 
consistent with recent published data. Predictions for future beauty mesons measurements have been included.
Ratios of strange to non-strange prompt charm mesons 
in central Pb + Pb and minimum bias (MB) $ p + p$ collisions 
at 2.76 TeV are also discussed. Minimum bias  $p + p$ collisions which 
constitute theoretical baseline in our calculations 
are studied at the centre of mass energies $\sqrt{s}$ = 2.76 TeV and 7 TeV.

\end{abstract}


\maketitle


\section{Introduction}

The phase transition from hadronic to
partonic degrees of freedom in ultra-relativistic nuclear collisions 
is a central focus of experiments at the CERN Large Hadron Collider (LHC)
\cite{qgp4,Muller:2012zq,Jacak:2012dx,Bazavov:2011nk}.
Heavy-flavor quarks are an ideal probe to study early dynamics 
in these nuclear collisions. 
Several theoretical studies predict a substantial enhancement of open 
charm production, associated with the formation of a plasma of deconfined 
parton matter relative to the case of a purely hadronic scenario 
without plasma formation 
\cite{Muller:1992xn,Shuryak:1992wc,Geiger:1993py,Kuznetsova:2006hx,He:2011qa}.
For reviews of heavy-flavor production in heavy-ion collisions 
see Refs.~\cite{Frawley:2008kk,Gale:2012hq,Mischke:2013sp,Andronic:2013awa}.
The study of open charm production allows one to 
probe the mechanisms of heavy-quark propagation, energy loss 
and hadronization in the hot dense medium formed in high-energy 
nucleus-nucleus collisions 
\cite{Mischke:2013sp,Andronic:2013awa,Dainese:2011vb,Dainese:2012ae,delValle:2012qw}. Heavy quarks are key observables in the study   
of thermalization of the initially created hot 
nuclear matter \cite{Zhu:2006er,Linnyk:2008hp}. 

Owing to their large mass, heavy quarks are produced predominantly in the 
initial phase of the collision via gluonic fusion processes
\cite{Lin:1994xma} and therefore probe the complete space-time evolution of 
the quark gluon plasma (QGP) matter. 
Their production rates are expected to be well described 
by perturbative Quantum Cromodynamics (pQCD)
at Fixed Order plus Next to-leading Logarithms (FONLL) 
\cite{Mangano:1991jk,cacciari_05,Cacciari:2012ny}.
 Measurements at RHIC energies  
\cite{Abelev:2006db,Zhang:2008kr,Adler:2005xv,Adare:2006nq} have shown that 
the gluon fusion process could also
dominates in heavy-ion collisions and that thermal processes might 
contribute later at low transverse momentum \cite{Uphoff:2010sh}.

The production and propagation of hard probes in nucleus-nucleus ($A + A$) 
collisions can be quantified by the nuclear modification factor (NMF)
\begin{equation}
\label{eq:RAA}
 R_{AA}\,(p_T) =\,\frac{(1/N_{\rm evt}^{AA})d^{2}\,N_{AA}/d^{2}\,p_{T}dy}
           {N_{\rm coll}(1/N_{\rm evt}^{pp})d^{2}\,N_{pp}/d^{2}\,p_{T}dy}
\end{equation} 
where, N$_{\rm evt}$ is the number of events and N$_{\rm coll}$
is the average number of binary nucleon-nucleon ($NN$) collisions,
and $d^{2}\,N/d^{2}\,p_{T}dy$ stand for the transverse 
momentum ($p_T$) and rapidity ($y$) differential yield of 
an observable measured in $A+A$ or proton-proton ($p+p$) collisions.
A value $R_{AA}\,(p_T) \neq 1$  would indicate contributions 
from initial and final-state effects.  These observables provide stringent 
constraints on theoretical predictions, in particular 
jet quenching in $A + A$ collisions at 
Relativistic Heavy Ion Collider (RHIC) and at Large Hadron 
Collider (LHC) energies. 

One of the most exciting discoveries at RHIC, 
was that heavy quark is suppressed by an amount similar to that of light quarks, for transverse momentum 
$p_T > 5 $ GeV/{\it c} \cite{Tannenbaum:2013wkn} (open charm RHIC puzzle).
This result was a surprize; it appears to disfavour the energy loss
explanation of suppression \cite{Gyulassy:1990ye,Baier:2000mf} 
based on the fact that heavy quarks should 
radiate much less than light quarks or gluons. In addition, the dead-cone 
effect \cite{Dokshitzer:2001zm} and other mechanisms 
\cite{Armesto:2005iq,Wicks:2007am}
are expected to introduce a mass-dependence in the coupling of 
hard partons with the medium constituents.
A possible solution to this puzzle \cite{Tannenbaum:2013wkn,Zichichi:2008zz}
 is  based on the assumption that in the standard model, 
Higgs Boson, which gives
mass to the electro-weak vector bosons, does not necessarily gives mass
to fermions and it can not be excluded that in a QCD coloured world,
all six quarks are nearly massless.    


The non-perturbative particle creation mechanisms in strong external fields
has a wide range of application not only in original 
 $e^{+}e^{-}$ pair creation on QED problems \cite{schwinger}, but also for pair 
creation (fermions and bosons) in strong non-Abelian electromagnetic fields 
\cite{Biro:1984cf,Gyulassy:1986jq,Tanji:2008ku,Ruffini:2009hg,Labun:2008re,Cardoso:2011cs,Nayak:2005pf,Nayak:2005yv,Levai:2008wf,Levai:2009mn,Levai:2011zza,Levai:2011zz}.
In a high-energy heavy-ion collision, strong color fields are expected 
to be produced between the partons of the projectile and target.
Theoretical descriptions of particle production in high energy $p+p$ and
$A +A$ collisions are based on the introduction of chromoelectric flux tube
({\em strings}) models \cite{Andersson:1983ia,Wang:1991hta}.
 String breaking picture \cite{Andersson:1983ia} 
is a good example of how to convert the kinetic 
energy of a collisions into field energy. 
Therefore, Schwinger mechanism is assumed to be an important mechanism for 
hadronic production. 
For a uniform chromoelectric flux tube with field (E) the probability 
to create a pair of quarks with mass (m), effective charge (e$_{\rm eff}$ = e/3),
and transverse momentum ($p_T$) per unit time and per unit volume 
is given by \cite{nussinov_80} :
\begin{equation}
P(p_T)\,d^2p_T 
=-\frac{|e_{\rm eff}E|}{4 \pi^3} ln \Bigg\{ 1-exp\left[-\frac{\pi(m^2+p_T^2)}
{|e_{\rm eff}E|} \right] \Bigg\} \,\,d^2p_T
\label{eq:Pt}
\end{equation}
The integrated probability ($P_m$) reproduces the classical Schwinger
results \cite{schwinger}, derived in spinor quantum
electrodynamics (QED) for $e^{+}e^{-}$ production rate,
when the leading term in Eq.~\ref{eq:Pt} is taken into account, {\it i. e.}:
\begin{equation}
P_m=\frac{(e_{\rm eff}E)^2}{4 \pi^3} \sum_{n=1}^{\infty}
\frac{1}{n^2}exp\left(-\frac{\pi\,m^2n}{|e_{\rm eff}E|}\right)
\label{eq:scwinger}
\end{equation}

In a string fragmentation phenomenology, it has been proposed
that the observed strong enhancement of strange particle
production in nuclear collisions
could be naturally explained via strong longitudinal 
color field effects (SLCF)~\cite{Gyulassy:1986jq}.
Recently, an extension
of Color Glass Condensate (CGC) theory has proposed a more detailed
dynamical ``GLASMA'' model \cite{larry_2009,McLerran:2008qi,Kharzeev:2006zm} 
of color ropes.
In the string models, strong longitudinal fields 
(flux tubes, effective strings) decay into new ones by 
 quark anti-quark ($q\bar{q}$ ) or 
diquark anti-diquark (qq-$\overline{\rm qq}$) pair 
production and subsequently hadronize to produce the 
observed hadrons. Due to confinement, the color 
of these strings is restricted to a small area in transverse space
\cite{Cardoso:2011cs}.
With increasing energy of the colliding particles, the number of
strings grows and they start to overlap, forming clusters. 
This can introduce a possible dependence of particle production on the
energy density \cite{Braun:2012kn}.


Heavy Ion Jet Interacting ({\small HIJING}) type models 
such as {\small HIJING1.0} \cite{Wang:1991hta},
 {\small HIJING2.0} \cite{Deng:2010mv,Deng:2010xg}
and {\small HIJING/B\=B} v2.0 
\cite{Pop:2004dq,Pop:2005uz,ToporPop:2007hb,Pop:2009sd,ToporPop:2010qz,ToporPop:2011wk,Barnafoldi:2011px,Pop:2012ug,Albacete:2013ei},
have been developed to study hadron productions 
in $p+p$, $p+A$ and $A+A$ collisions.   
These models are based on a two-component geometrical 
model of mini-jet production and soft interaction and has incorporated 
nuclear effects such as {\em shadowing} (nuclear modification of the parton 
distribution functions) and {\em jet quenching}, via final state jet
medium interaction.  
In the {\small HIJING/B\=B} v2.0 model \cite{ToporPop:2007hb,ToporPop:2010qz}
we introduced new dynamical effects associated with
long range coherent fields ({\it i.e}, strong longitudinal color fields, SCF),
including baryon junctions and loops \cite{Pop:2005uz,ripka_lnp2004}.  
At RHIC energies we have shown  \cite{Pop:2004dq,Pop:2005uz,ToporPop:2007hb}
that the dynamics of strangeness production 
deviates considerably from calculations based on Schwinger-like
estimates for homogeneous and constant color fields \cite{schwinger}, 
and points to the possible 
contribution of fluctuations of transient strong color fields (SCF).
These fields are similar to those which could appear in a 
 {\em glasma} \cite{McLerran:2008qi} at initial stage of the collisions.
In a scenario with QGP phase
transitions the typical field strength of SCF at RHIC energies 
was estimated to be about 5-12 GeV/fm \cite{Magas:2002ge}.

The tunneling process mechanism of heavy Q\=Q pair 
production have been revisited \cite{Cohen:2008wz} and
pair production in time-dependent
electric fields have been studied \cite{Hebenstreit:2008ae}.
It is concluded that particles with large momentum are likely to have 
been created earlier than particle with small momentum and in addition 
during a very short period  $\Delta \tau$  
($\Delta \tau \approx 10 {\rm t}_{\rm Q}$, where the Compton time 
${\rm t}_{\rm Q}=1/{\rm m}_{\rm Q}$) the  standard Schwinger formula 
({\it i.e.} with a constant electric field), strongly
underestimates the particle number density.

In a previous paper \cite{Pop:2009sd} effects of strong longitudinal color 
electric fields (SCF) on the open charm production in nucleus-nucleus
({\it A} + {\it A}) collisions at RHIC energies were investigated 
within the framework of the {\small HIJING/B\=B v2.0} model
\cite{Pop:2004dq,Pop:2005uz,ToporPop:2007hb}.
It was shown that a three fold increase of the effective string tension
results in a sizeable enhancement ($ \approx $ 60-70 \%) 
of the total open charm production cross sections
($\sigma^{\rm NN}_{{\it c}\,\bar{{\it c}}}$) in comparison with the results
obtained without SCF effects. 
At top LHC energy ($\sqrt{s} = 14$ TeV) the  {\small HIJING/B\=B v2.0} model predicts an increase in $p+p$ collisions
of $\sigma^{\rm NN}_{{\it c}\,\bar{{\it c}}}$ by approximately an order of magnitude
 \cite{Pop:2009sd}. 
Moreover, in this work we offer an alternative
explanation of the open charm RHIC puzzle since
the calculated nuclear modification factor of $D^0$ mesons shows 
at moderate transverse momentum (${\rm p}_{\rm T}$)
a suppression consistent with RHIC data 
\cite{Abelev:2006db,Zhang:2008kr,Adler:2005xv,Adare:2006nq}.  
String fusion and percolation effects on heavy flavour production 
have also been  
discussed in Refs.~\cite{Merino:2009py,Pajares:2010vm} at RHIC and LHC energies.
The production patern for heavy quarks in both of these 
non-perturbative approaches
becomes similar to that of the light quarks via the 
Schwinger mechanism \cite{schwinger} and result on an expected enhancement 
of heavy quark pairs Q\=Q.

Recently, the total open charm cross sections were   
reported in $p + p$ collisions at $\sqrt{s} = 2.76$ and $\sqrt{s} = 7$ 
TeV by ALICE \cite{Abelev:2012vra,ALICE:2011aa,Abelev:2012tca}
 , ATLAS \cite{ATLAS:2011fea,Rossi:2011xp,ATLAS:2012ama}
and LHCb \cite{Aaij:2013mga} Collaborations.
Measurements of open-heavy flavor $p_T$ differential production 
cross sections ($ \sigma_{\rm inel} d^{2}\,N_{AA}/d^{2}\,p_{T}dy$)  
in Pb + Pb Collisions  at a center of mass energy per nucleon pair 
$\sqrt{s_{\rm NN}}=2.76 $ TeV have also been published by 
the ALICE Collaboration \cite{Dainese:2012ae,delValle:2012qw}, 
\cite{ALICE:2012ab,Bailhache:2012gr,Grelli:2012yv,ValenciaPalomo:2012jd}.

In $p+p$ collisions at $\sqrt{s} = 7$ TeV the $p_T$-differential 
production cross sections of prompt charmed mesons 
(D$^0$, D$^+$, D$^{*+}$, D${_s}{^+}$) at mid-rapidity ($|y| < 0.5$) are  
compatible with the upper limit 
of the FONLL predictions \cite{Maciula:2013wg},
leaving room for possible new dynamical mechanisms. 
Note that, the models with different parametrization of un-integrated gluon distributions (UGDF) 
significantly underpredict the experimental data \cite{Maciula:2013wg}.
In contrast the models with general-mass variable-flavor-number scheme (GM-VFNS)
overpredict data \cite{Kniehl:2012ti}.

RHIC results show that heavy quark lose energy in the medium,
but a possible quark-mass hierarchy 
predicted in Ref.~\cite{Armesto:2005iq} has not been established,
{\it i.e.}, a smaller suppression  expected when going from 
the mostly gluon-originated light flavour hadrons ({\it e.g.}, pions)
to D and B mesons \cite{ALICE:2012ab}.  
At LHC energies, prompt D mesons present a similar suppression as 
charged particles and this observation is challenging for most
theoretical and phenomenological analysis 
\cite{Dainese:2012ae,delValle:2012qw}. The model calculations 
for nuclear modification factors of prompt charmed mesons 
in Pb + Pb collisions at $\sqrt{s_{\rm NN}}=2.76 $ TeV indicate a reasonable 
agreement with data
\cite{Sharma:2009hn,He:2011pd,Horowitz:2011cv,Horowitz:2011wm,Alberico:2011zy,
Monteno:2011gq,Gossiaux:2009mk,Gossiaux:2010yx,Buzzatti:2011vt}
but only for moderate and high transverse momentum ($p_T$), 
{\it i.e.} $p_T > $ 5 GeV/c, where the suppression is 
a factor of 2.5-4 in comparison with binary scaling  \cite{ALICE:2012ab}.
However, the description at low transverse momentum ($p_T \leq $ 4 GeV/{\it c})
is more challenging for the currently available theoretical model calculations.
The expected $p$ + Pb collisions data will provide new valuable information on
possible initial-state effects in the low-momentum region.

The {\small HIJING/B\=B v2.0} model has successfully 
described the global observables and identified particle (ID) data,  
including (multi)strange particles production in  {\it p}+{\it p} 
\cite{ToporPop:2010qz,Pop:2012ug}
{\it p} + Pb \cite{Barnafoldi:2011px,Albacete:2013ei}
and Pb + Pb collisions \cite{ToporPop:2011wk} at RHIC and LHC energies.
In this paper we extend our study 
to open prompt charm mesons production (D$^0$, D$^+$, D$^{*+}$, D${_s}{^+}$) 
as measurements have been recently published \cite{Abelev:2012vra,ALICE:2011aa,Abelev:2012tca}. The setup and input parameters 
used here are taken from previous works 
(see Refs.~\cite{ToporPop:2011wk,Pop:2012ug,Albacete:2013ei}). 
We explore dynamical effects associated with
long range coherent fields ({\it i.e.} strong color fields, SCF),
including baryon junctions and loops, with emphasis 
on the novel open charm observables measured at LHC energies
in {\it p}+{\it p} collisions at  $\sqrt{s} = 2.76$ and 
 $\sqrt{s} = 7$ TeV. 
The nuclear final state effects (jet quenching) and initial state effects
(shadowing) are discussed in term of nuclear modification factors $R_{\rm AA}(p_T)$ 
measured in Pb + Pb collisions at $\sqrt{s_{\rm NN}} = 2.76$ TeV
\cite{ALICE:2012ab,Bailhache:2012gr,Grelli:2012yv,ValenciaPalomo:2012jd}. 
In addition, in order to better identify initial state effects, 
predictions for nuclear modification factor $R_{pA}(p_T)$ in {\it p} + Pb collisions at $\sqrt{s_{\rm NN}} = 5.02$ TeV are also presented.

\section{Outline of {\small HIJING/B\=B v2.0} model. Setup and input.}

\subsection{Strong color field. String tension.}

In this paper we present the results of calculations for
different observables measured in $p+p$, $p$ + Pb and Pb + Pb collisions
at LHC energies. Therefore, we consider useful to the reader 
to include a summary of the main input parameters which have been
determined in Refs.~\cite{ToporPop:2011wk,Pop:2012ug,Albacete:2013ei}
and that are used in the present analysis.
This is the subject of Sec. II where we describe the basics phenomenology 
embedded in the {\small HIJING/B\=B} v2.0 model.

Based on the assumption that Higgs Boson, which gives
mass to the electro-weak vector bosons, does not necessarily gives mass
to fermions and that in a QCD coloured world,
all six quarks are nearly massless \cite{Zichichi:2008zz},
we investigate if the Schwinger mechanism could play a role in the 
non-perturbative soft production 
of heavy quarks (Q = c, or b), within the framework of the 
 {\small HIJING/B\=B} model.
For a uniform chromoelectric flux tube with field ({\it E}),
for a heavy quark pair ($Q\bar{Q}$) 
the production rate per unit volume
is given by~\cite{Gyulassy:1986jq,Cohen:2008wz,Hebenstreit:2008ae} 

\begin{equation}
\Gamma =\frac{\kappa^2}{4 \pi^3} 
{\text {exp}}\left(-\frac{\pi\,m_{Q}^2}{\kappa}\right),
\label{eq:Gamma}
\end{equation}
Note that $\Gamma$ is given by the first term in the series
of integrated probability P$_m$ (Eq.~\ref{eq:scwinger}). 
For $Q = c $ (charm) or $Q = b$ (bottom) , $m_{\rm Q}=1.27,\;{\rm or}\; 4.16$ GeV (with $\pm 1\%$ uncertainty \cite{Steinhauser:2008pm}), 
and $\kappa=|e_{\rm eff}E|$ is the effective string tension.
For a color rope, if the  {\em effective} string tension value ($\kappa$) 
increases from vacuum value 
$\kappa$ = $\kappa_{0}$ = 1.0 GeV/fm to an in medium value 
$\kappa $ = 3.0 GeV/fm,
the pair production rate per unit volume for charm pairs 
would increase from $\approx 1.4\,\cdot 10^{-12}$ to 
$\approx 3.5\,\cdot 10^{-4}$ fm$^{-4}$.
This can lead to  a net soft tunneling production
comparable to the initial hard FONLL pQCD prediction.
In the {\small HIJING/B\=B} model (which is a two component model) 
the string/rope fragmentation is the only soft source of multiparticle
production and multiple minijets provide a semi-hard additional
source that is computable within collinear factorized standard pQCD
with initial and final radiation (DGLAP evolution \cite{parisi_77}).

A measurable rate for spontaneous pair production requires 
strong chromoelectric fields, such 
that $\kappa/m_{\rm Q}^2\,\,>$ 1 some of the time.
Introducing strong longitudinal electric field within string models,
result in a highly suppressed production rate of heavy $Q\bar{Q}$ pair 
($\gamma_{Q\bar{Q}}$) related to light quark pairs 
($q\bar{q}$). From Eq.~\ref{eq:Gamma}
one obtain  \cite{Cohen:2008wz} the suppression factor $\gamma_{Q\bar{Q}}$ 
\begin{equation}
\gamma_{Q\bar{Q}} = \frac{\Gamma_{Q\bar{Q}}}{\Gamma_{q\bar{q}}} =
{\text {exp}} \left(-\frac{\pi(m_{Q}^2-m_q^2)}{\kappa} \right), 
\label{eq:gammaQ}
\end{equation}
The suppression factors are calculated for 
$Q={\rm qq}$ (diquark), $Q = s$ (strange), $Q = c$ (charm), or $Q = b$ (bottom) ($q=u,\,d$ stand for light quarks).

The current quark masses are $m_{\rm qq}$ = 0.45 GeV \cite{ripka:2005},  
$m_s=0.12$ GeV, $m_{\rm c}=1.27$ GeV, and $m_b=4.16$ GeV \cite{pdg:2010}.
The constituent quark masses of light non-strange quarks 
are $M_{u,d}$ = 0.23 GeV, of the strange quark is $M_s$=0.35 GeV 
\cite{armesto2001}, and of the diquark is $M_{\rm qq}=0.55 \pm 0.05$ GeV 
\cite{ripka:2005}.
In our calculations, we use $M^{\rm eff}_{qq}$ = 0.5 GeV, 
$M^{\rm eff}_{s}$ = 0.28 GeV, $M^{\rm eff}_{c}$ = 1.27 GeV.
Therefore, for the vacuum string tension value $\kappa_0$ = 1 GeV/fm, the above formula from Eq.~\ref{eq:gammaQ} results  \cite{Pop:2012ug} in a suppression of heavier quark production according to
$u$ : $d$ : ${\rm qq}$ : $s$ : $c$ $\approx$ 1 : 1 : 0.02 : 0.3 : 10$^{-11}$. 
For a color rope, on the other hand,
if the effective string tension value $\kappa$ 
increases to  $\kappa = f_{\kappa} \kappa_0$ (with $ f_{\kappa} > 1$)
the value of $\gamma_{Q\bar{Q}}$ increases.
Equivalently, a similar increase of $\gamma_{Q\bar{Q}}$ could be obtained by
a decrease of quark masses from $m_{Q}$ to  $m_{Q}/\sqrt{f_{\kappa}}$.
We have shown that this dynamical mechanism improves considerably the 
description of the strange meson/hyperon 
data at the Tevatron and at LHC energies \cite{ToporPop:2010qz}.

At ultra-high energy, $A + A$ collisions can also be described  
as two colliding sheets of Color Glass Condensate (CGC). 
In the framework of this model
it was shown that in early stage of collisions a strong longitudinal 
color-electric field is created \cite{larry_2009}.
Saturation physics is based on the observation
that small-x hadronic and nuclear wave functions, and, thus the 
scattering cross sections as well, are described by the same internal 
momentum scale known as the saturation scale, $Q_{\rm sat}$.
In  $p+p$ collisions at LHC energies the saturation scale is proportional to the charged particle density at midrapidity, 
$ Q_{\rm sat,p}^2(s) \propto (dN_{\rm ch}/d\eta)_{\eta =0} $.
An analysis of $p+p$ data up to $\sqrt{s}$ = 7 TeV
has shown that, with  the $k_T$ factorized gluon fusion approximation
\cite{Gribov:1984tu},
the growth of the charged particle density at midrapidity can be accounted for
if the saturation scale grows with c. m. energy ($\sqrt{s}$) as
\cite{McLerran:2010ex}:
\begin{equation}
Q_{\rm sat,p}^2(s) = Q_{0p}^2 (s/s_0)^{\lambda_{\rm CGC}},
\label{eq:larry10}
\end{equation}
with $\lambda_{\rm CGC} \approx 0.11$.
It has been argued that  
the saturation scale increases with atomic number in  nucleus-nucleus
collisions. A natural way 
is to assume that $Q_{\rm sat,A}^2$ is 
proportional to the number of participants in the collisions, {\it i.e.}, as 
$Q_{\rm sat,A}^2 \propto Q_{\rm sat,p}^2(s) A^{1/3}$ \cite{Kharzeev:2006zm}.
It has been proposed that the gluonic partons saturating the heavy ion 
collisions produce in the CGC perturbative flux tubes with an original width 
of transverse size, of the order of $1/Q_{\rm sat,A}$ 
\cite{Dumitru:2008wn,Dusling:2009ni}, flux
tubes persisting during the evolution of quark gluon plasma.

In the Lund hadronization model \cite{Andersson:1983ia,Wang:1991hta},
the large number of particle produced in heavy ion collisions are reproduced
with string fragmentation. 
When a pair of QCD charge and anti-charge are pulled apart, a flux tube
of fields develops between the pair. Flux tubes, approximated by a thin string
for modelling, are extended and non-linear objects. They have been 
observed in Latice QCD \cite{Cardoso:2011cs}.
The flux tubes utilized to simulate $A+A$ collisions may have a 
string tension almost one order of magnitude larger than the 
fundamental string tension linking a mesonic quark-antiquark pair 
\cite{Biro:1984cf,Cardoso:2011cs}. 

The initial energy densities ($\epsilon_{\rm ini}$) are computed from 
the square of the field components
\cite{Cardoso:2011cs}. Within our phenomenology
$\epsilon_{\rm ini}$ is proportional to mean field values $<E^2>$,
and using the relation $\kappa= e_{\rm eff} E$, means 
$\epsilon_{\rm ini} \propto \kappa^2$.
Here, we do not take into consideration the energy corresponding to 
color magnetic field, which are one order of magnitude smaller that 
those corresponding to SCF \cite{Cardoso:2011cs}). 
Using Bjorken relation the $\epsilon_{\rm ini}$ is proportional with 
charged particle density at midrapidity , and thus  
$ \kappa^2 \propto (dN_{\rm ch}/d\eta)_{\eta =0} $.
A similarity with the phenomenology embedded in the CGC model
is obvious, and we obtain $\kappa \propto Q_{sat,p}$ as 
discussed in Ref.~\cite{ToporPop:2010qz}.
In  Ref.~\cite{ToporPop:2010qz}, to describe the energy dependence 
of the charged particle density at mid rapidity in $p+p$ collisions
up to the LHC energies, we use a power law dependence   
\begin{equation} 
\kappa(s)= \kappa_{0} \,\,(s/s_{0})^{0.06}\,\,{\rm GeV/fm},
\label{eq:kappa_sup_old}
\end{equation}
consistent (within the error) with that 
deduced in CGC model \cite{McLerran:2010ex}.

We have shown in Ref.~\cite{Pop:2012ug} 
that combined effects of hard and soft sources of 
multiparticle production as embedded in the {\small HIJNG/B\=B} v2.0 model
can reproduce charged particle density at midrapidity 
and identified particle spectra (including (multi)strange particles)  
 in $p+p$ collisions in the range  $0.02<\sqrt{s}< 7$ TeV, 
by an energy dependent string tension $\kappa (s)$,
with a somewaht reduced power law :
\begin{equation}
\kappa(s)= \kappa_{0} \,\,(s/s_{0})^{0.04}\,\,{\rm GeV/fm},
\label{eq:kappa_sup}
\end{equation}

This new parametrization from Eq.~\ref{eq:kappa_sup} does not affect 
significantly the entropy embedded in the model 
and the charged particle densities 
at midrapidity are also well described (see Ref.~\cite{Pop:2012ug}).
Equation~\ref{eq:kappa_sup} leads to an increasing value for the mean 
string tension from 
  $\kappa$ = 1.5 GeV/fm at $\sqrt{s}$ = 0.2 TeV (top RHIC energy) 
to  $\kappa$ = 2.0 GeV/fm at $\sqrt{s}$ = 7 TeV.
The sensitivity of the calculations to 
string tension values ($\kappa$) for different observables 
have been studied in previous papers \cite{Pop:2005uz,ToporPop:2007hb,Pop:2009sd,ToporPop:2010qz,ToporPop:2011wk,Pop:2012ug}. 

This constitute the only modification of the model parameters discussed in
our previous paper~\cite{Pop:2012ug}. 
Our phenomenological parametrizations Eq.~\ref{eq:kappa_sup},
is strongly supported  by data on the square root of charged particle densities at midrapidity ($\sqrt{(dN_{\rm ch}/d\eta)_{\eta =0}}$). 
Within the error the  $\sqrt{(dN_{\rm ch}/d\eta)_{\eta =0}}$ show a
power law dependence proportional to $s^{0.05 }$
for inelastic $p+p$ interactions  and to  $s^{0.055}$ 
for non-single diffractive events~\cite{Aamodt:2010pb,ALICE:2012xs}.

In $A + A$ collisions the effective string tension value 
could also increase due to in-medium effects~\cite{ToporPop:2011wk},
or possible dependence on number of participants. 
This increase is also quantified in our phenomenology by an analogy 
with CGC model. 
We consider for the mean value of the string tension an energy and 
mass dependence, 
$\kappa(s,A) \propto Q_{\rm sat,A} (s,A) \propto  Q_{\rm sat,p} (s) A^{1/6}$.
Therefore, for $A + A$ collisions we use in the presnt analysis,
a power law dependence $\kappa$ = $\kappa(s,A)$
\begin{equation}
\kappa(s,A)_{\text LHC} = \kappa(s) A^{0.167}
= \kappa_{0} \,\,(s/s_{0})^{0.04} A^{0.167}\,\,{\rm GeV/fm},
\label{eq:kapsA} 
\end{equation}

Equation~\ref{eq:kapsA} leads to $\kappa(s,A)_{\text LHC} \approx 5$ GeV/fm, 
in Pb +Pb collisions 
at c.m. energy per nucleon $\sqrt{s_{NN}}$ = 2.76 TeV.  
First heavy-ion data at the LHC, {\it i.e.}, 
charged particle density and nuclear modification factor $R_{\rm PbPb}$  
are only slightly different (see Sec IIIB, Fig.~\ref{fig:fig2}) 
from those calculated in Ref.~\cite{ToporPop:2011wk} where 
a higher value of $\kappa$,   
$\kappa(s,A)_{\text LHC} =\kappa_{0} \,\,(s/s_{0})^{0.06} A^{0.167} \approx 6$ GeV/fm was used. 
The reason for this small effect is that 
the suppression factors $\gamma_{Q\bar{Q}}$,  
approach unity in Pb + Pb collisions at $\sqrt{s_{NN}}$ = 2.76 TeV, 
for the string tension values $\kappa \geq 5$ GeV/fm.

The mean values of the string tension 
$\kappa(s)$ for $p + p$ collisions (Eq.~\ref{eq:kappa_sup}) 
and $\kappa(s,A)$ for $A + A$ collisions (Eq.~\ref{eq:kapsA})
are used in the present calculations. These lead
to a related increase of the various suppression factors,
as well as an enhancement of the intrinsic (primordial) 
transverse momentum $k_T$.
These include:  i) the ratio of production rates of  
diquark-quark to quark pairs (diquark-quark suppression factor),  
$\gamma_{{\rm qq}} = \Gamma({\rm qq}\overline{{\rm qq}})/\Gamma(q\bar{q})$;
ii) the ratio of production rates of strange 
to non-strange quark pairs (strangeness suppression factor), 
$\gamma_{s} = \Gamma(s\bar{s})/\Gamma(q\bar{q})$;
iii) the extra suppression associated with a diquark containing a
strange quark compared to
the normal suppression of strange quark ($\gamma_s$),
$\gamma_{us} = (\Gamma({\rm us}\overline{{\rm us}})/\Gamma({\rm
  ud}\overline{{\rm ud}}))/(\gamma_s)$;
iv) the suppression of spin 1 diquarks relative to spin 0 ones
(in addition to the factor of 3 enhancement of the former based on
counting the number of spin states), $\gamma_{10}$; and 
v) the (anti)quark ($\sigma_{q}'' = \sqrt{\kappa/\kappa_0} \cdot \sigma_{q}$)
and  (anti)diquark ($\sigma_{\rm {qq}}'' = \sqrt{\kappa/\kappa_0}
\cdot {\it f} \cdot \sigma_{{\rm qq}}$) Gaussian  
width of primordial (intrinsic) transverse momentum $k_T$.
In the above formulae for $\sigma_{q}''$ and $\sigma_{\rm {qq}}''$ we use 
 $\sigma_{q}$ = $\sigma_{\rm qq}$ = 0.350 GeV/{\it c} as default value 
(in absence of SCF effects) for Gaussian width of quark (diquark)  
intrinsic transverse momentum distribution.

Moreover, to better describe the baryon/meson anomaly seen in data
at RHIC and LHC energies, a specific implementation of J\=J loops, had to be 
introduced (for details see Refs.~\cite{ToporPop:2011wk,Pop:2012ug}).
The absolute yield of charged particles, $dN_{\rm ch}/d\eta$ is also sensitive
to the low $p_T < \, 2$ GeV/{\it c} nonperturbative hadronization
dynamics that is performed via LUND \cite{Andersson:1986gw} 
string JETSET \cite{Bengtsson:1987kr} fragmentation as
constrained from lower energy $e+e, e+p, p+p$ data.  
The conventional hard pQCD mechanisms are calculated 
 in {\small HIJING/B\=B v2.0} via the {\small PYTHIA}
\cite{Sjostrand:2006za} subroutines. 
The advantage of {\small HIJING/B\=B v2.0}  over {\small PYTHIA}
is the ability to include novel SCF color rope effects 
that arise from longitudinal fields amplified by the random walk 
in color space of the high x valence partons in {\it A}+{\it A} collisions. 
This random walk could induce   
a very broad fluctuation spectrum of the effective string tension.

In the present work we will study only the effect 
of a larger effective value $\kappa > $ 1 GeV/fm
on the production of prompt charmed mesons (D$^0$, D$^+$, D$^{*+}$, D${_s}{^+}$)
measured in Pb + Pb and $p+p$ collisions at LHC energies.
The model is based on the time-independent strength of color field while in 
reality the production of $Q\bar{Q}$ pairs is a far-from-equilibrium,
time and space dependent complex phenomenon.
Therefore, we can not investigate in details possible fluctuations
which could appear due to these more complex dependences.

\subsection{Nuclear shadowing and Jet quenching.}

As mentioned above, in HIJING the string/rope fragmentation is not the only 
{\em soft source} of multiparticle
production and multiple minijets provide a semi-hard additional
source that is computable within collinear factorized standard pQCD
with initial and final radiation (DGLAP evolution \cite{parisi_77}).  
Within the HIJING model, one assumes that nucleon-nucleon
collisions at high energy can be divided into {\em soft} and
{\em hard} processes with at least one pair of jet with
transverse momentum, $p_{T}>p_0 $. A cut-off (or saturation) 
scale $p_0$ in the final jet production has to be introduced 
below which the high density of initial interactions
leads to a non-perturbative mechanisms which in the HIJING framework
is characterized by a finite soft parton cross section $\sigma_{\rm soft}$.
The inclusive jet cross section $\sigma_{\rm jet}$ at leading order
(LO) \cite{Eichten:1984eu} is 

\begin{equation}
\label{eq:sigma_jet}
 \sigma_{jet}=\int_{p_0^2}^{s/4}\mathrm{d}p_T^2\mathrm{d}y_1\mathrm{d}y_2 
 \frac{1}{2}\frac{\mathrm{d}\sigma_{jet}}{\mathrm{d}p_T^2
 \mathrm{d}y_1 \mathrm{d}y_2},
\end{equation}
where,
\begin{equation}
\label{eq:dif_sigma_jet}
 \frac{\mathrm{d}\sigma_{jet}} {\mathrm{d}p_T^2 \mathrm{d}y_1 
\mathrm{d}y_2} =K \sum_{a,b}x_1f_a(x_1,p_T^2)x_2f_b(x_2,p_T^2) 
\frac{\mathrm{d}\sigma^{ab}(\hat{s},\hat{t},\hat{u})}{\mathrm{d}\hat{t}}
\end{equation}
depends on the parton-parton cross section $\sigma^{ab}$ and parton 
distribution functions (PDF), $f_a(x,p_T^2)$. The summation runs over all 
parton species; $y_1$ and $y_2$ are the rapidities of the scattered
partons; $x_1$ and $x_2$ are the light-cone momentum 
fractions carried by the initial partons.
The multiplicative $K$ factor ($K \approx 1.5-2$) account for    
the next-to-leading order (NLO) corrections to the leading order (LO)
jet cross section \cite{Eskola:1995cj,Campbell:2006wx}. 
In the default {\small HIJING} model \cite{Wang:1991hta,Wang:1991us}, the  
Duke-Owens parameterization \cite{Duke:1983gd} for PDFs of nucleons is used.
With the Duke-Owens parameterization for PDFs, an energy independent
cut-off scale $p_0=$2 GeV/$c$ and a constant soft
parton cross section $\sigma_{soft}=57$ mb are sufficient to
reproduce the experimental data on total and inelastic cross sections 
and the hadron central rapidity density in $p+p(\bar{p})$ collisions
 \cite{Wang:1991hta,Wang:1991us}.   

The largest uncertainty in mini-jet cross sections is the strong 
dependence on the minimum transverse momentum scale cut-off, $p_0$. 
In this paper the results for $p+p$ collisions are obtained using 
the same set of parameters for hard scatterings as in the default  
{\small HIJING} model \cite{Wang:1991us}. 
Using a constant momentum cut-off $p_0$ = 2 GeV/c 
 in central $A+A$ collisions, 
the total number of minijets per unit transverse area 
for independent multiple jet production,
 could exceed the limit \cite{Deng:2010mv,Deng:2010xg}
\begin{equation}
\label{eq:taa_limit}
\frac{T_{AA}(b) \sigma_{\rm jet}}{\pi R_{A}^{2}} \leq 
\frac{p_0^2}{\pi},
\end{equation}  
where $T_{AA}(b)$ is the overlap function of
$A+A$ collisions and $\pi/p_0^2$ is the intrinsic 
transverse size of a minijet with transverse momentum $p_0$.
Therefore, an increased value of $p_0$ with c.m. energy 
per nucleon $ \sqrt{s_{\rm NN} }$
is required by the experimental data indicating that 
the {\em coherent interaction} becomes important. 
Moreover, we have to consider an energy and nuclear size dependent 
cut-off $p_0(s,A)$, in order to ensure the applicability
of the two-component model for $A+A$ collisions.
It was shown~\cite{ToporPop:2011wk} that  
the pseudorapidity distribution of charged particle in central 
nucleus-nucleus collisions at RHIC and LHC 
energies can be well described
if we consider a scaling law of the type $C A^{\alpha} \sqrt{s}^{\beta}$
\begin{equation}
p_0(s,A)= 0.416\,A^{0.128}\,\sqrt{s}^{\,\,0.191}\,\,{\rm GeV}/c
\label{eq:p0}
\end{equation}
A similar dependence was used in pQCD + Saturation model to 
predict global obsevables at LHC energies~\cite{Eskola:2001vs}. 
The main difference is the value of the proportionality constant 
( $C_{\rm HIJ}$ = 0.416 vs. $C_{\rm esk}$ = 0.208).
The value $C_{\rm esk}$ = 0.208 used in Ref.~\cite{Eskola:2001vs,Eskola:1999fc}
results in an overestimate of the charged 
particle density by a factor of approximately two at LHC energies.
These effective values are not expected to be valid for peripheral 
$A+A$ or for $p+p$ collisions.

The above limit for incoherent mini-jet production should in fact also depend 
on impact-parameter~\cite{Eskola:2000xq}.  Such dependence is not
included in the present calculations.
Instead, in the  HIJING model an impact-parameter dependence 
of the gluon shadowing is considered in the parameterization of the 
parton shadowing factor $S_{a/A}$  (see below).  
Due to shadowing effects the observed A-exponent ($\alpha=0.128$) 
in Eq.~\ref{eq:p0} is somewhat less than the 
number expected in the saturated scaling limit ($p_0(s,A) \sim A^{1/6}$)
\cite{Eskola:1999fc}. 

One of the main uncertainty in calculating 
charged particle multiplicity density 
in Pb + Pb collisions is the nuclear modification of parton
distribution functions, especially gluon distributions at small $x$.
In {\small HIJING}-type models, one assumes that 
the parton distributions in a nucleus (with atomic number A and 
charge number Z), $f_{a/A}(x,Q^2)$, are 
factorizable into parton distributions of nucleons ($f_{a/N}$)
and the parton(a) shadowing factor ($S_{a/A}$),
\begin{equation} 
f_{a/A}(x,Q^2) = S_{a/A}(x,Q^2)Af_{a/N}(x,Q^2)
\end{equation}
 We assume that the shadowing effect for gluons and quarks is the same, 
and neglect also the QCD evolution ($Q^2$ dependence of the shadowing effect).
At this stage, the experimental data unfortunately 
can not fully determine the $A$ dependence of the shadowing effect. 
We follow the $A$ dependence as proposed in 
Ref.~\cite{Wang:1991hta} and use the following parametrization,
\begin{eqnarray}
        S_{a/A}(x)&\equiv&\frac{f_{a/A}(x)}{Af_{a/N}(x)} \nonumber\\
         &=&1+1.19\log^{1/6}\! A\,[x^3-1.2x^2+0.21x]\nonumber\\
 & &-s_{a}(A^{1/3}-1)[1 -\frac{10.8}{{\rm log}(A+1)}\sqrt{x}]e^{-x^2/0.01},
          \label{eq:shadow}\\
        s_a &=&0.1,\label{eq:shadow1}
\end{eqnarray}

The term proportional to $s_a$ in 
Eq.~\ref{eq:shadow} determines the shadowing for $x < x_0 = 0.1$, 
and has the most important nuclear dependence, while the rest gives the overall 
nuclear effect on the structure function in $x > x_0$ with some very slow 
$A$ dependence. This parametrization can fit the overall nuclear 
effect on the quark structure function in the small and medium 
$x$ region \cite{Wang:1991hta}. 
Because the remaining term of Eq.~\ref{eq:shadow} has a very slow $A$ 
dependence, we consider only the impact parameter dependence 
of $s_a$. In fact most of the jet production occurs in 
the small $x$ region where shadowing is important:
\begin{equation}
        s_a(b)=s_a\frac{5}{3}(1-b^2/R_A^2)
                        \label{eq:rshadow}
\end{equation}
In the above equation $R_A$ is the radius of the nucleus, and the factor $s_a$ is taken the same for quark and for gluon $s_a = s_q = s_g = 0.1$ . 

The LHC data indicate that such quark (gluon) shadowing is
required to fit the centrality dependence of the central charged
particle multiplicity density in Pb + Pb collisions \cite{ToporPop:2011wk}. 
This constraint on quark (gluon) shadowing is indirect and model dependent.
Therefore, it will be important to study  
quark(gluon) shadowing in $p+A$ collisions at the LHC.
In contrast, in {\small HIJING2.0}~\cite{Deng:2010mv,Deng:2010xg}, a 
different {\it A} parametrization ($(A^{1/3}-1)^{0.6}$) and
much stronger impact
parameter dependence of the gluon ($s_g=0.22-0.23$) 
and quark ($s_q=0.1$) shadowing factor is used in order 
to fit the LHC data. 
Because of this stronger gluon shadowing the jet quenching effect is
neglected~\cite{Deng:2010mv}.
Note, all {\small HIJING}-type models assume a scale-independent form 
of shadowing parametrization (fixed $Q^2$). This approximation could   
break down at very large scale due to the dominance of gluon emission
dictated by the DGLAP \cite{parisi_77} evolution equation.
The default {\small HIJING1.0} parametrization of the fixed $Q_0^2=2$ GeV$^2$ 
shadowing function \cite{Wang:1991hta} leads to substantial reduction at the LHC of the global multiplicity in $p$ + Pb and Pb + Pb collisions. 
It is important to emphasize that the
 {\em no shadowing} results are substantially reduced in 
{\small HIJING/B\=B}2.0~\cite{ToporPop:2011wk,Barnafoldi:2011px,Albacete:2013ei}, relative to the {\em no shadowing} predictions within
 {\small HIJING/1.0} from Ref.~{\protect\cite{Wang:1991hta}},
 because both the default minijet cut-off $p_0=2 $ GeV/{\it c}
and the default vacuum string tension $\kappa_0=1 $ GeV/fm 
(used in {\small HIJING1.0}) are generalized
to vary monotonically with centre of mass (c.m.) energy $\sqrt{s}$ 
and atomic number, $A$. 

As discussed above, systematics of $p+p$ 
and Pb+Pb multiparticle production from RHIC to the LHC 
are used to fix the energy ($\sqrt{s}$) and the $A$ dependence of the
cut-off parameter 
$p_0(s,A) = 0.416  \; \sqrt{s}^{0.191} \; A^{0.128}$ GeV/{\it c} 
and a mean value of the string tension 
$\kappa(s,A) = \kappa_0\;(s/s_0)^{0.04}\;A^{0.167}$ GeV/fm in $A+A$ collisions
\cite{Pop:2012ug}.
For $p$ + Pb collisions at $\sqrt{s_{\rm NN}}$ = 5.02 TeV, 
the above formulae lead to $p_0 = 3.1$ GeV/{\it c} (calculated as a mean value
of $p_0^{PbPb} = 4.2 $ GeV/{\it c} and $p_0^{pp} = 2 $ GeV/{\it c}).
The measurement of  initial energy density produced in  $p$ + Pb collisions
would help us to determine better the effective  
value of string tension, $\kappa$ in $p$ + Pb collisions. Therefore, 
in the present calculations we consider 
$\kappa_{pPb} = 2.1$ GeV/fm at  $\sqrt{s_{\rm NN}}$ = 5.02 TeV, which fit 
charged particle production ($dN_{\rm ch}/d\eta$) \cite{ALICE:2012xs,ALICE:2012mj} 
at mid-pseudorapidity in minimum bias events selection 
of $p$ + Pb interactions \cite{Albacete:2013ei}.  
For $p+p$ collisions at  $\sqrt{s}$ = 5.02 TeV we use 
a constant cut-off parameter $p_{0pp} = 2 $ GeV/{\it c} and an effective string 
tension value of $\kappa_{pp} = 1.9 $ GeV/fm.


The ALICE Collaboration at the LHC published first experimental 
data on the charged 
hadron multiplicity density at mid-rapidity in central (0-5\%) Pb + Pb 
collisions at $\sqrt{s_{\rm NN}}$ = 2.76 TeV \cite{Aamodt:2010pb,Aamodt:2010cz}.
In this experiment the collaboration confirmed the presence of 
jet quenching ($R_{AA} < 1$)  \cite{Aamodt:2010jd,Abelev:2012hxa}.
These results provide stringent constraints on the theoretical predictions 
in Pb + Pb collisions at LHC energies.

In order to describe new Pb + Pb data 
\cite{Aamodt:2010pb,Aamodt:2010cz,Aamodt:2010jd,Abelev:2012hxa}, we modified in 
{\small HIJING/B\=B v2.0} model (see~Ref.~\cite{ToporPop:2011wk}) 
the main parameters describing hard partons interactions.
For a parton $a$, the energy loss per unit distance can be expressed as   
$dE_a/dx = \epsilon_a/\lambda_a$,
where $\epsilon_a$ is the radiative energy loss per scattering and 
$\lambda_a$ is the mean free path (mfp) of the inelastic scattering.
For a {\em quark jet} at the top RHIC energy ($\sqrt{s_{\rm NN}}$ = 0.2 TeV)
$(dE_q/dx)_{\rm RHIC} = 1$ GeV/fm and mfp $(\lambda_q)_{\rm RHIC} = 2$ fm 
\cite{ToporPop:2007hb}. The initial parton density is 
proportional to the final hadron multiplicity density.
The charged particle density at mid-pseudorapidity at $\sqrt{s_{\rm NN}}$ = 2.76 TeV is
a factor of 2.2 higher than  at $\sqrt{s_{\rm NN}}$ = 0.2 TeV~\cite{Aamodt:2010pb}. 
Therefore, for a {\em quark jet} at the LHC  the energy loss (mfp) should increase
(decrease) by a factor of $\approx 2.0$ and become 
$(dE_q/dx)_{\rm LHC} \approx 2$ GeV/fm and mfp $(\lambda_q)_{\rm LHC} \approx 1$ fm. For a gluon jet $dE_g/dx$ = 2 $dE_q/dx$.
Throughout this analysis we will consider the results 
with the following set of parameters for hard interactions: 
{\it i.e.}, $K = 1.5$; $dE_q/dx = 2$ GeV/fm; $\lambda_q=1 $ fm.  
Since there is always a coronal region with an average length $\lambda_q$
in the system where the produced parton jets will escape without scattering or 
energy loss, the suppression factor can never be infinitely small.
For the same reason, the suppression factor should also depend on $\lambda_q$.
It is difficult to extract information on both $dE_q/dx$ and $\lambda_q$
simultaneously from the measured spectra in a model independent way~\cite{Wang:1998bha}.

In the next section we show that a constant radiative energy loss mechanism 
(dE/dx=const) and jet quenching mechanism as implemented in 
the {\small HIJING/B\=B v2.0} model provides a good description of  
suppression at intermediate and high $p_T$ ($4< p_T < 15$ GeV/{\it c})
for charged particles and prompt charmed mesons production in Pb+Pb collisions 
at LHC energies.

\section{Numerical Results and Discussion}

\subsection{Open prompt charm production in $p+p$ collisions}

\begin{figure} [h!]
\centering
\includegraphics[width=\linewidth,height=8.0 cm]{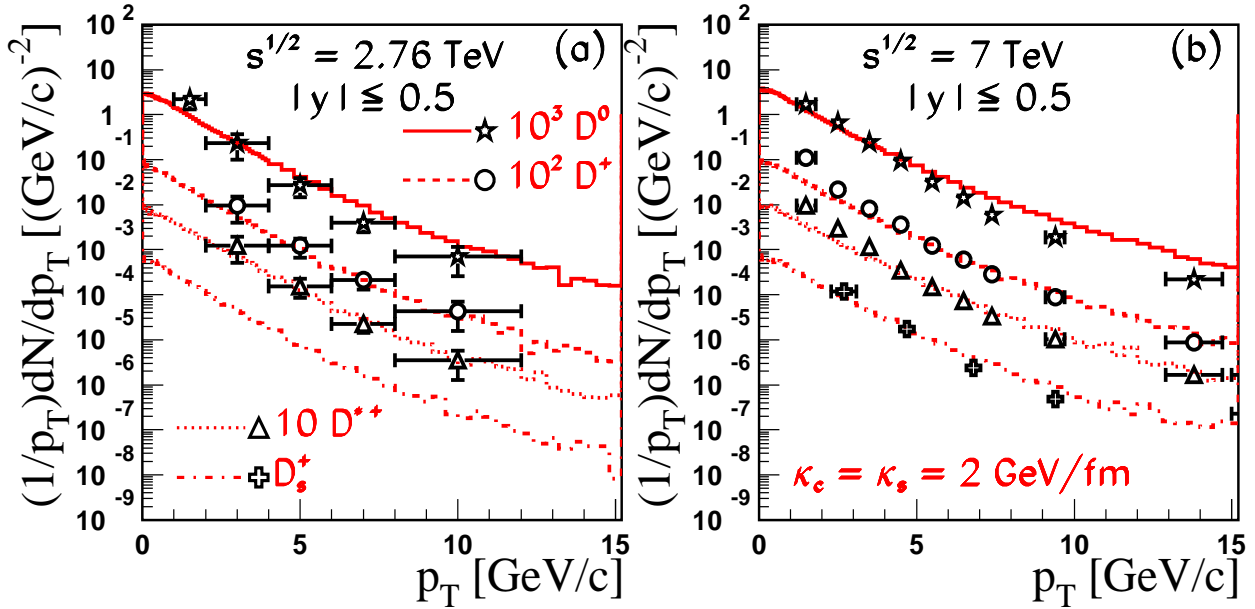}
\vskip 0.5cm
\caption[pp at 2.76 and 7TeV pt distributions] 
{\small (Color online) {\small HIJING/B\=B v2.0} predictions for
$p_T$ distributions at mid-rapidity for $p + p \rightarrow $ (D + \=D)/2 + X  with
D = D$^0$ (solid histograms); D = D$^{+}$ (dashed histogram); 
D = D$^{*+}$ (dotted histograms); and D = D${_s}^{+}$ (dash-dotted histogram).
The results are compared to data at  $\sqrt{s}$ = 2.76 TeV (left panel) 
from Ref.~\cite{Abelev:2012vra} 
and at $\sqrt{s}$ = 7 TeV (right panel) 
from Refs.~\cite{ALICE:2011aa,Abelev:2012tca}. 
For clarity, the experimental data and theoretical 
results are multiplied with a factor indicated in the figure.
Only statistical error bars are shown.
\label{fig:fig1}
}
\end{figure}

The ALICE Collaboration has reported measurements of the 
transverse momentum distribution
of open prompt charmed mesons (D$^0$, D$^+$, D$^{*+}$, D${_s}{^+}$) in $p + p$
collisions at $\sqrt{s}$ = 7 TeV \cite{ALICE:2011aa,Abelev:2012tca},
and  of (D$^0$, D$^+$, D$^{*+}$) at $\sqrt{s}$ = 2.76 TeV \cite{Abelev:2012vra}
in the central rapidity range $|y| \le 0.5 $.
{\em Prompt} indicates D mesons produced at the $p+p$ interaction point, 
either directly in the hadronization of the charm quark 
or in strong decays of excited charm resonances. 
The contribution from weak decays of beauty mesons, which 
give rise to feed-down D mesons, were subtracted.
The model calculations include SCF effects as discussed in Section II A.
The energy dependence  of string tension from Eq.~\ref{eq:kappa_sup}, 
$\kappa(s)= \kappa_{0} \,\,(s/s_{0})^{0.04}\,\,{\rm GeV/fm}$, predict 
a modest increase when going from $\sqrt{s}$ = 2.76 TeV 
($\kappa$ = 1.88 GeV/fm) to  $\sqrt{s}$ = 7 TeV ($\kappa$ = 2.03 GeV/fm). 
Therefore, to calculate open prompt charmed mesons  
production we consider 
the same value of average string tension for charm and strange quark, {\it i.e},
$\kappa_c $ = $\kappa_s \,$ $\approx$ 2 GeV/fm. 
The theoretical results are compared to data in Fig.~\ref{fig:fig1}.
Predictions for  D${_s}{^+}$ meson at  $\sqrt{s}$ = 2.76 TeV 
are also included. The agreement between theory and experiment 
is good within experimental uncertainties, 
except at  $\sqrt{s}$ = 7 TeV where while the average cross section is well reproduced
the predicted spectrum has  a somewhat shallower slope than the data.

The results at  $\sqrt{s}$ = 7 TeV 
are also reasonably well described by FONLL calculations
\cite{Cacciari:2012ny}, NLO pQCD calculations \cite{Mangano:1991jk},
and GM-VFNS model for $p_T > 3 $ GeV/{\it c} \cite{Kniehl:2012ti}.
The limited statistics of the experimental data at $\sqrt{s}$ = 2.76 TeV 
\cite{Abelev:2012vra}
prevents the use of these measurements 
as a baseline for $R_{\rm PbPb}$ studies  of 
prompt charmed hadrons. Instead in Refs.~  \cite{delValle:2012qw,Abelev:2012vra}
in calculating  $R_{\rm PbPb}$ at $\sqrt{s_{\rm NN}}$ = 2.76 TeV 
the baseline $p + p$ spectrum was obtained by a pQCD-driven $s$-scaling
of the $p + p$ differential cross section from  $\sqrt{s}$ = 7 TeV 
to $\sqrt{s}$ = 2.76 TeV \cite{Cacciari:2012ny,delValle:2012qw}. 
The scaled D meson cross sections at 2.76 TeV were found to be 
consistent with those measured with only a limited precision 
of 20-25 \% \cite{Abelev:2012vra}.
In this paper we use as baseline for calculations of NMF 
$R_{\rm PbPb}$ at $\sqrt{s_{\rm NN}}$ = 2.76 TeV, $p + p$ theoretical results 
obtained within {\small HIJING/B\=B v2.0} model.

\subsection{Nuclear modification factors in Pb + Pb collisions 
at $\sqrt{s_{\rm NN}} $ = 2.76 TeV}

\begin{figure} [h!]
\centering

\includegraphics[width=\linewidth,height=8.0cm]{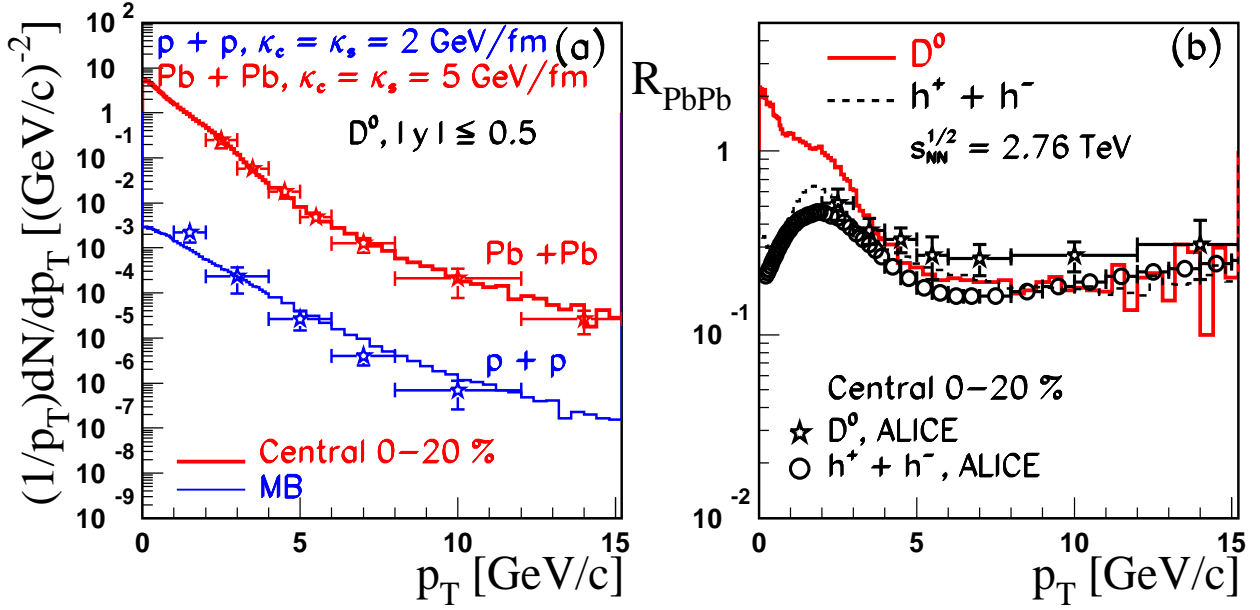}
\vskip 0.5cm
\caption[R_AA of D0 for Pb+Pb relative to pp] 
{\small (Color online) 
Part (a): {\small HIJING/B\=B v2.0} predictions for $p_T$ distributions at 
mid-rapidity for Pb + Pb $\rightarrow$ 
(D$^0$ + \=D$^0$)/2 + X (upper histogram), 
and for $p+p$ collisions (lower histogram).
Part (b): The $p_T$ dependence of NMF ${\rm R}_{\rm AA}({\rm p}_{\rm T})$ 
for D$^0$ mesons (solid histogram)  
and charged particles (dashed histogram) in central (0-20 \%) Pb + Pb collisions.
Data are from ALICE Collaboration for D$^0$ (stars) \cite{ALICE:2012ab} and 
for charged particles (open circles) \cite{Abelev:2012hxa}.
Error bars include only statistical uncertainties.  
\label{fig:fig2}}
\end{figure}

The nuclear modification factor $R_{\rm PbPb}$ has been measured
by the ALICE Collaboration 
for the centrality classes 0-20 \% and 40-80 \% in Pb + Pb collisions
at $\sqrt{s_{\rm NN}}$ = 2.76 TeV 
for prompt D$^0$, D$^+$ and D$^{*+}$ ~\cite{ALICE:2012ab}.
The results of the {\small HIJING/B\=B v2.0} model for $p_T$ spectra 
in $p+p$ (lower histogram) and central 0-20 \% Pb + Pb collisions (upper histogram) 
are compared to data  \cite{ALICE:2012ab} in Fig.~\ref{fig:fig2} (left panel).
For Pb + Pb collisions the results include quenching and shadowing effects
as discussed in Sec. II B.  
In the calculations we take into account the variation of strong color (electric) field with energy and 
the size of the colliding system. The assumed average string tension is  
$\kappa_c$ =$\kappa_s$ = 2.0 GeV/fm and  $\kappa_c$ = $\kappa_s$ = 5.0 GeV/fm
 for {\it p} + {\it p} and Pb + Pb collisions, respectively.
The agreement with the data is good except perhaps for 
$p+p$ reactions, where the slope of the predicted spectrum is a bit shallower
than that seen in the data, as it was already mentioned in the Sec. III A.

The transverse momentum spectra of identified particles carrying light quarks and their 
azimuthal distributions are well described by hydrodynamical models 
\cite{Heinz:2013th,Gale:2013da} at low $p_T$.
The calculated spectra for ${\it D}^0$-mesons show a small shoulder at very low ${\rm p}_{\rm T}$ 
indicating possible infuence of the radial flow. 
However, as far as in the string model the pressure is not considered
it is not expected to describe the sizable 
elliptic flow of heavy quarks as observed by the ALICE Collaboration~\cite{ALICE:2012ab}.

The transverse momentum dependence of the D$^0$ nuclear modification factor 
 $R_{\rm PbPb}^{D^0}$ is shown in Fig.~\ref{fig:fig2} (right panel).  
At transverse momentum $p_T > 6$ GeV/{\it c} the charmed mesons show a suppression 
factor of $\approx$ 4. Also shown is a comparison with results for 
lighter quark species, specifically charged hadrons \cite{Abelev:2012hxa}.
{\small HIJING/B\=B} model calculations have shown \cite{ToporPop:2011wk} 
that the charged-pions $R_{\rm PbPb}^{\pi}$
coincides with that of charged hadrons above $p_T \approx 6$ GeV/{\it c} and are lower 
by 25 \% -30 \% in the $p_T$ range 2-4 GeV/c.
At high $p_T > $ 6 GeV/{\it c} the calculated D$^0$ meson suppression is comparable with 
that of charged particles (and $\pi$ mesons) within experimental uncertainties.
This result indicates that the energy loss of charm quarks 
is rather similar with the one of lighter quarks or gluons, in contrast with 
previous theoretical studies \cite{Dokshitzer:2001zm,Wicks:2007am}.

\begin{figure} [h!]
\centering

\includegraphics[width=\linewidth,height=8.0cm]{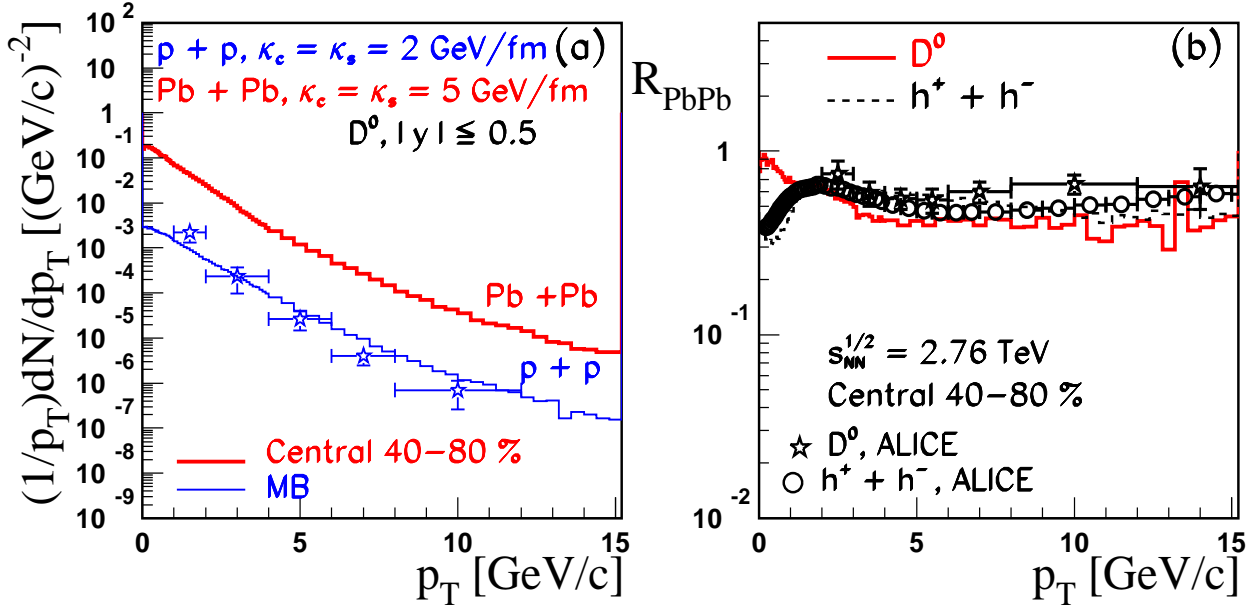}
\vskip 0.5cm
\caption[R_AA of D0 for Pb+Pb relative to pp c40-80] 
{\small (Color online) Comparison of {\small HIJING/B\=B v2.0} predictions of
$p_T$ distributions (left panel) and 
NMF ${\rm R}_{\rm AA}({\rm p}_{\rm T})$ for D$^0$  
and charged particles in semi-peripheral
 (40-80 \%) Pb + Pb collisions (right panel). The histograms have the same meaning as in Fig.~\ref{fig:fig2}.
Data are from ALICE Collaboration for D$^0$ (stars) \cite{ALICE:2012ab} and 
for charged particles (open circles) \cite{Abelev:2012hxa}.
Error bars include only the statistical uncertainties. 
\label{fig:fig3}
}
\end{figure}

At low ${\rm p}_{\rm T}$ ($0\,<\,{\rm p}_{\rm T}\,< 4 $ GeV/c), 
the non-perturbative production mechanism via SCF  
produces  a difference between D$^0$ and charged particles (mainly $\pi$ mesons).
The reason for this difference is that yields of charged particles are reduced due to
conservation of energy \cite{Pop:2005uz} and  
yields of D-mesons are enhanced due to an increase 
of $s\,\bar{s}$ and $c\,\bar{c}$ pair production (see Eq.~\ref{eq:gammaQ}).
In this range of $p_T$, the model predicts a quark-mass hierarchy, 
{\it i.e.}, $R_{\rm PbPb}^{\pi} \, < \,  R_{\rm PbPb}^{ch} \, < \,R_{\rm PbPb}^{D}$.
Within model phenomenology we can interpret the above result as evidence for 
``{\em in-medium mass modification}'' of charm quark, due to 
possible chiral symmetry restoration \cite{Kharzeev:2005iz}.
An in-medium mass modification has also been predicted 
near the phase transition ({\it i.e.}, at lower energy) in \cite{Tolos:2005ft}.
In contrast, the statistical hadronization model \cite{Andronic:2008gm}
predicts no medium  effects at RHIC and LHC energies. 
Preliminary recent ALICE data \cite{delValle:2012qw,Grelli:2012yv} 
suggest a decrease in going from low to high $p_T$ albeit with big errors. 
Measurements with good statistics at low $p_T$ 
are needed in order to draw a definite conclusion concerning
the shape of the transverse momentum dependence of  $R_{\rm PbPb}^{D} (p_T) $.
Similar results (not included here) are obtained for
prompt D$^+$ and D$^{*+}$ meson production.

When compare with Fig.\ref{fig:fig2}, ~Fig.~\ref{fig:fig3} shows
 that the {\small HIJING/B\=B} model predict less  
suppression for D$^0$ meson (solid histogram)
from $\approx 4$ to $\approx 1.6$ 
when going from from central 0-20 \% to semi-peripheral 40-80 \% 
Pb + Pb collisions.  
Once more, at high $p_T > $  6 GeV/{\it c} the D$^0$ meson suppression is comparable with 
those of charged particles (dashed histogram) within experimental uncertainties.
These results are consistent with data for D$^0$ meson
\cite{ALICE:2012ab} and for charged particles~\cite{Abelev:2012hxa}.
At low $p_T$ the split between  D$^0$ meson and charged particles 
is considerably
reduced except at very low  $p_T$ ($p_T < $ 1 GeV/{\it c}) 
where a modest quark-mass hierarchy $R_{\rm PbPb}^{ch} \, < \,R_{\rm PbPb}^{D}$ is predicted.

\begin{figure} [h!]
\centering
\includegraphics[width=\linewidth,height=8.0cm]{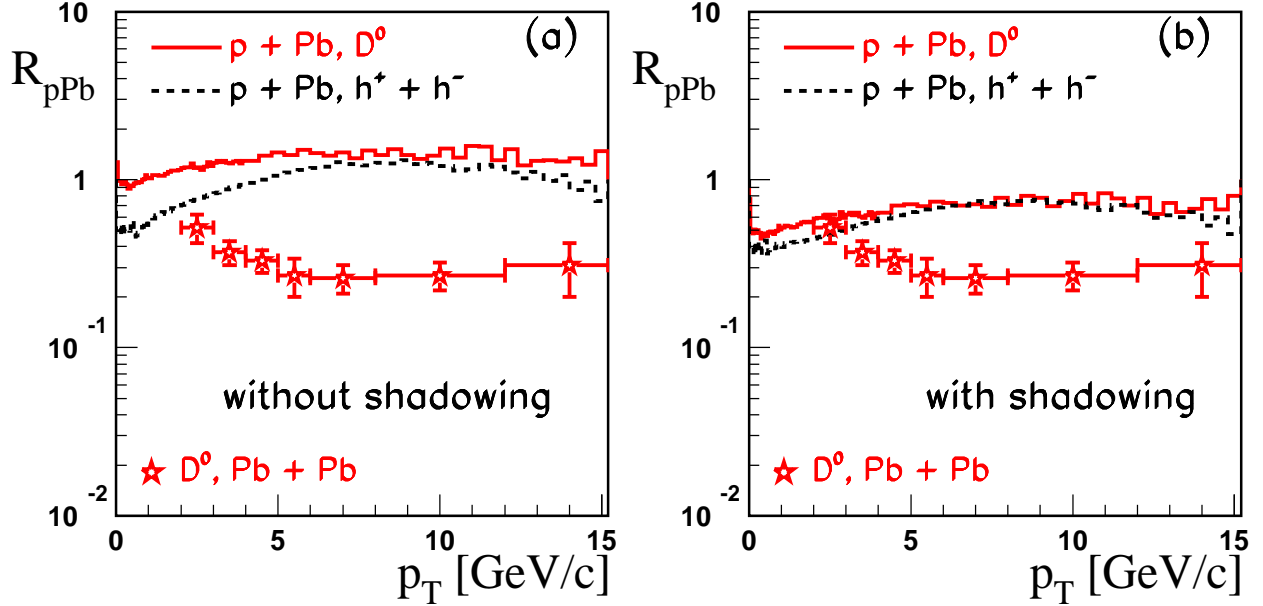}
\vskip 0.5cm\caption[pt spectra for D0, minimum-bias,central] 
{\small (Color online) The  {\small HIJING/B\=B v2.0} model predictions for
 $ R_{\rm pPb}$ of D$^{0}$ meson (solid histograms) and charged particles
(dashed histograms) in the 0-20 \% centrality class $p$ + Pb 
colisions at $ \sqrt{s_{\rm NN}} $ = 5.02 TeV. 
The results assuming no shadowing (left panel) and with shadowing
(right panel) are compared with experimental data on  $R_{\rm PbPb}$ 
for D$^0$ meson in the same centrality class (0-20 \%)
at $ \sqrt{s_{\rm NN}} $ = 2.76 TeV. The data are from Ref.~ \cite{ALICE:2012ab}.
Only statistical error bars are shown.
\label{fig:fig4}
}
\end{figure}

The suppression observed in NMF $R_{\rm PbPb}^{D^0} < 1$ 
has contributions from initial and final states. 
Initial state effects (such as nuclear shadowing and gluon saturation)
could be identified from the study  of open charm production 
in {\it p} + Pb collisions.
The initial production
of $c\,\bar{c}$ pairs by gluon fusion might be suppressed 
due to gluon shadowing.
We recall that shadowing is a depletion of the low-momentum parton
distribution in a nucleon embedded in a nucleus compared to a free
nucleon. In the kinematic range of interest the nuclear shadowing 
will reduce the PDF for partons with nucleon momentum fraction x 
below 10$^{-2}$. There is a considerable uncertainty 
(up to a factor of 3) in the amount of shadowing
predicted at RHIC and LHC energies by the different models with HIJING 
predicting the strongest effect \cite{Li:2001xa,d'Enterria:2008ge}.
The model predictions of $R_{\rm pPb}^{\rm D^{0}}$ in 
  $p$ + Pb collisions at $\sqrt{s_{NN}} $ = 5.02 TeV are presented 
for two scenarios, without (left panel) 
and with nuclear shadowing (right panel) in Fig.~\ref{fig:fig4},
and  compared to data of  $R_{\rm PbPb}^{D^0}$ obtained in the same centrality 
class at  $\sqrt{s_{NN}} $ = 2.76 TeV~\cite{ALICE:2012ab}.  
We use shadowing parameterizations as discussed in Sec.~II B.
Calculations without shadowing show no suppression except at low $p_T$
where one observes some difference between D$^0$ and charged particles.
Taking into account nuclear shadowing, the model predicts 
a suppression of $\approx 30 $ \% at high $p_T$ for both charged particles (dashed histogram) and D$^0$ mesons (solid histogram).
From this result, we may conclude that the strong 
suppression (a factor of $\approx$ 4) observed for  $R_{\rm PbPb}^{\rm D^{0}}$ \cite{ALICE:2012ab}
is a final state effect ({\it e.g.}, radiative and collisional energy loss in the QGP matter).
Note that for minimum bias measurements $R_{\rm pPb}^{\rm ch}$ is better described in a scenario without shadowing 
effects~\cite{Barnafoldi:2011px,Albacete:2013ei}.   
Since we expect higher sensitivity to shadowing effects for D$^0$ meson than for charged 
particles, measurements of $R_{\rm pPb}^{\rm D}$ at LHC energies  
could help to resolve this puzzle.

\begin{figure} [h!]
\centering

\includegraphics[width=\linewidth,height=8.0cm]{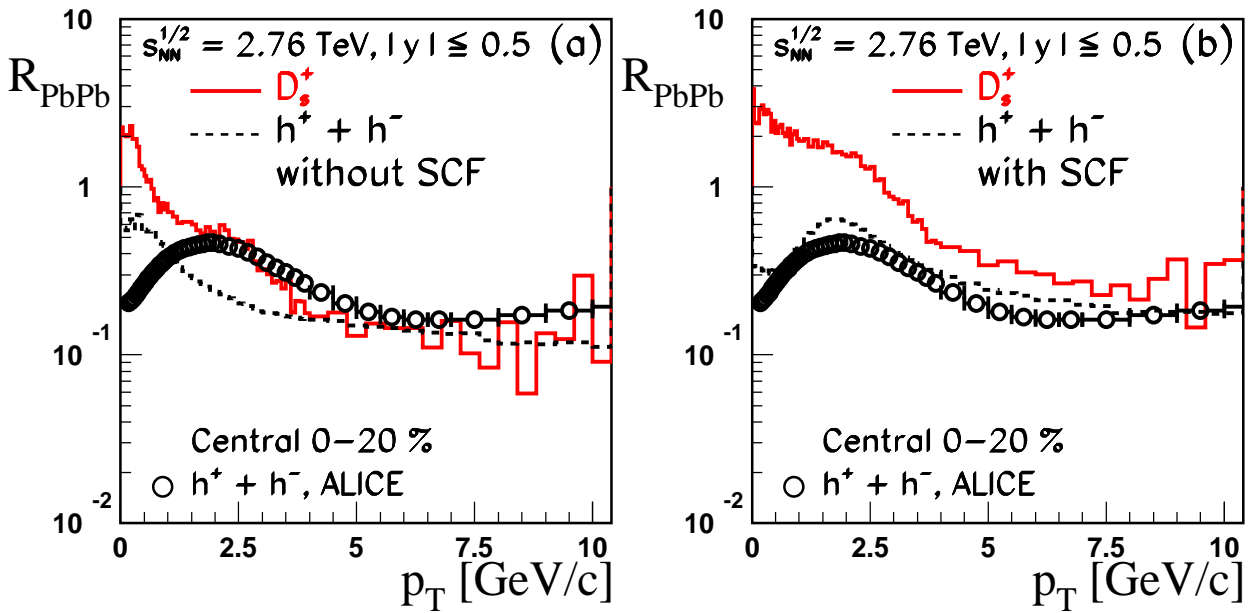}
\vskip 0.5cm\caption[R_AA of D_s for Pb + Pb relative to pp] {\small (Color online) 
Comparison of {\small HIJING/B\=B v2.0} predictions of
nuclear modification factor
${\rm R}_{\rm PbPb}({\rm p}_{\rm T})$ for $D_s^+$ (solid histograms) 
and charged particles (dashed histograms)  in central
(0-20 \%) Pb + Pb collisions at mid-rapidity. 
The results are presented for a scenario without SCF effects (left panel) and 
with SCF effects (right panel) (see text for details).
Data are from ALICE Collaboration \cite{Abelev:2012hxa}.
Error bars include only statistical uncertainties.
\label{fig:fig5}}
\end{figure}

Due to its strange quark content  
the study of production of prompt charmed mesons  D$_s^+$ ($c\,\bar{s}$) and  D$_s^-$ ($\bar{c}\,s$) is of particular interest. Our model predicts  
higher sensitivity to SCF effects for strange-charmed meson D$_s^+$ than for 
the non-strange charmed mesons (D$^0$, D$^+$, D$^{*+}$).  
In Fig.~\ref{fig:fig5} theoretical predictions for the $p_T $ dependence
of $R_{\rm PbPb}^{\rm D_s}$ for  D$_s^+$ mesons (solid histograms)
and  $R_{\rm PbPb}^{\rm ch}$ for charged particles (dashed histograms)
are presented for two scenarios: {\em without} (left panel) and {\em with}
(right panel) SCF effects. 
The calculations {\em without} SCF contributions assume 
for the string tension a vacuum value  
$\kappa_c$ = $\kappa_s$ = $\kappa_0 $ = 1 GeV/fm while the results {\em with} SCF 
are obtained including the energy and mass dependent, 
$\kappa_c$ = $\kappa_s$ $\approx$ 5 GeV/fm (see Sec. II B).
The calculations also include shadowing and quenching effects. 
The importance of in medium string tension values   
$\kappa_c$ = $\kappa_s$ = 5 GeV/fm is supported by the data. Only 
with SCF effects included, the model describes well charged particle NMF.
SCF induces a difference at low ${\rm p}_{\rm T}$ ($0\,<\,{\rm p}_{\rm T}\,< 4 $ GeV/c) between strange-charmed mesons D$_s^+$ and charged particles,
via non-perturbative production mechanism. 
The yields of strange-charmed mesons D$_s^+$ are enhanced due to an increase 
of $c\bar{c}$  and  $s\bar{s}$ pairs production (see Eq.~\ref{eq:gammaQ}).
In this range of $p_T$ the model predicts a quark-mass hierarchy, 
{\it i.e.}, $R_{\rm PbPb}^{\pi} \, < \,  R_{\rm PbPb}^{ch} \, < \,R_{\rm PbPb}^{D_s}$ ,
similar with those seen for non-strange charmed mesons.

The first preliminary experimental results of $R_{\rm PbPb}^{\rm D_{s}}$ for D$_s$ mesons 
in centrality class 0-7.5 \% Pb + Pb collisions at  $\sqrt{s_{NN}} $ = 2.76 TeV
\cite{Innocenti:2012ds} show at high $p_T$ 
a suppression factor of $\approx$ 5 and is compatible within 
uncertainties with those obtained for non-strange D-mesons.
However, at lower and moderate transverse momenta $ 2.5 < p_T < 8$ GeV/c 
the measured NMF $R_{\rm PbPb}^{\rm D_{s}}$   
\cite{Innocenti:2012ds} indicates values higher 
than the results shown in Fig.~\ref{fig:fig5} (right panel).  
We studied if one can find a scenario that would give a larger enhancement
of total yields for D$_s$ mesons. We consider the effect of    
a further increase of mean value of the string tension for charm quark from
$\kappa_c = 5$ GeV/fm to $\kappa_c = 10 $ GeV/fm, keeping a constant  
$\kappa_s = 5$ GeV/fm for strange quark. This allow to test a possible 
flavor dependence of $\kappa$, as suggested in Ref.~\cite{Levai:2009mn}.
These calculations (not included here) result in only a modest increase of 
  $R_{\rm PbPb}^{\rm D_{s}}$ by approximately 10-15 \%.  
For values of sting tension between 5 - 10 GeV/fm a saturation seems to  
set in, as an effect of energy and momentum conservation constraints. 

Due to large uncertainties in the data~\cite{Innocenti:2012ds} 
we can not draw yet a firm conclusions on possible enhancement of strange-charmed mesons over non-strange one as predicted by our approach.
Note that, at low and moderate $p_T$ ($ 0 < p_T < 8 $ GeV/{\it c}) 
other complex dynamical mechanisms 
such as transport, diffusion, and coalescence could play an important 
role in a description of  
the $R_{\rm PbPb}^{\rm D_{s}}$ for D$_s$ mesons at RHIC and LHC energies~\cite{He:2012df,Alberico:2013bza,Cao:2013ita}.
High statistics measurements in this $p_T$ range could help to disentangle  
between different approaches.

\begin{figure} [h!]
\centering
\includegraphics[width=\linewidth,height=8.0cm]{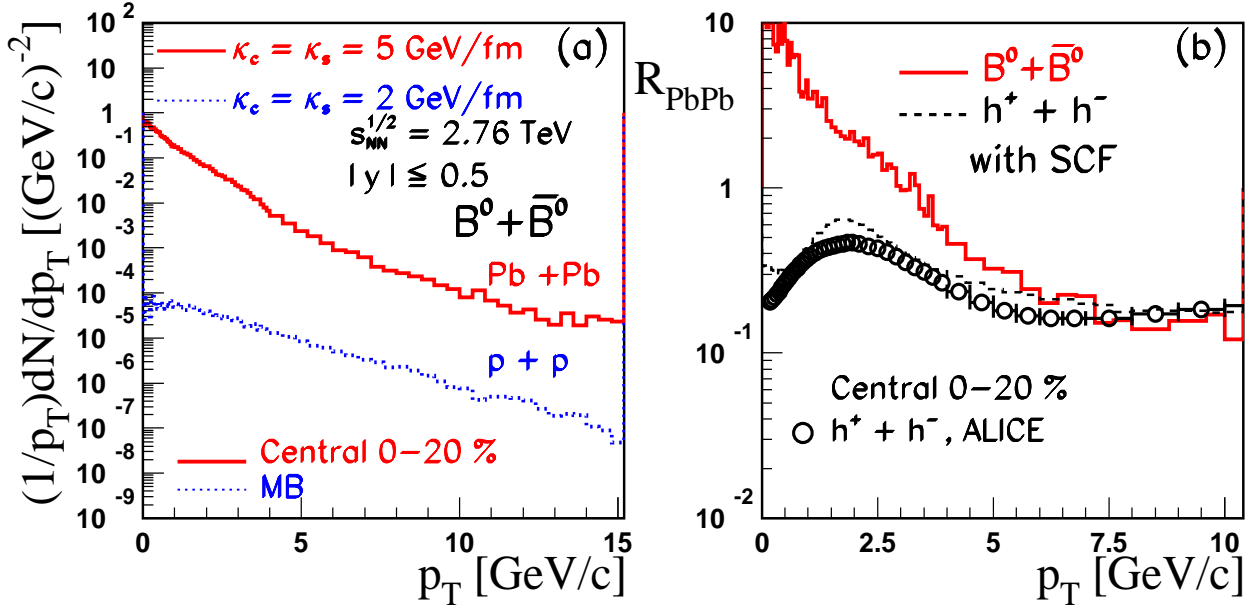}
\vskip 0.5cm
\caption[R_AA of B0 for Pb+Pb relative to pp] 
{\small (Color online) 
Part (a): {\small HIJING/B\=B v2.0} predictions for $p_T$ distributions at 
mid-rapidity for Pb + Pb $\rightarrow$ 
(B$^0$ + \=B$^0$)/2 + X (upper histogram), 
and for $p+p$ collisions (lower histogram).
Part (b): The $p_T$ dependence of NMF ${\rm R}_{\rm AA}({\rm p}_{\rm T})$ 
for B$^0$ mesons (solid histogram)  
and charged particles (dashed histogram) in central (0-20 \%) Pb + Pb collisions. Data are from ALICE Collaboration 
for charged particles (open circles) \cite{Abelev:2012hxa}.
Error bars include only statistical uncertainties.  
\label{fig:fig6}}
\end{figure}

\begin{figure} [h!]
\centering
\includegraphics[width=\linewidth,height=8.0cm]{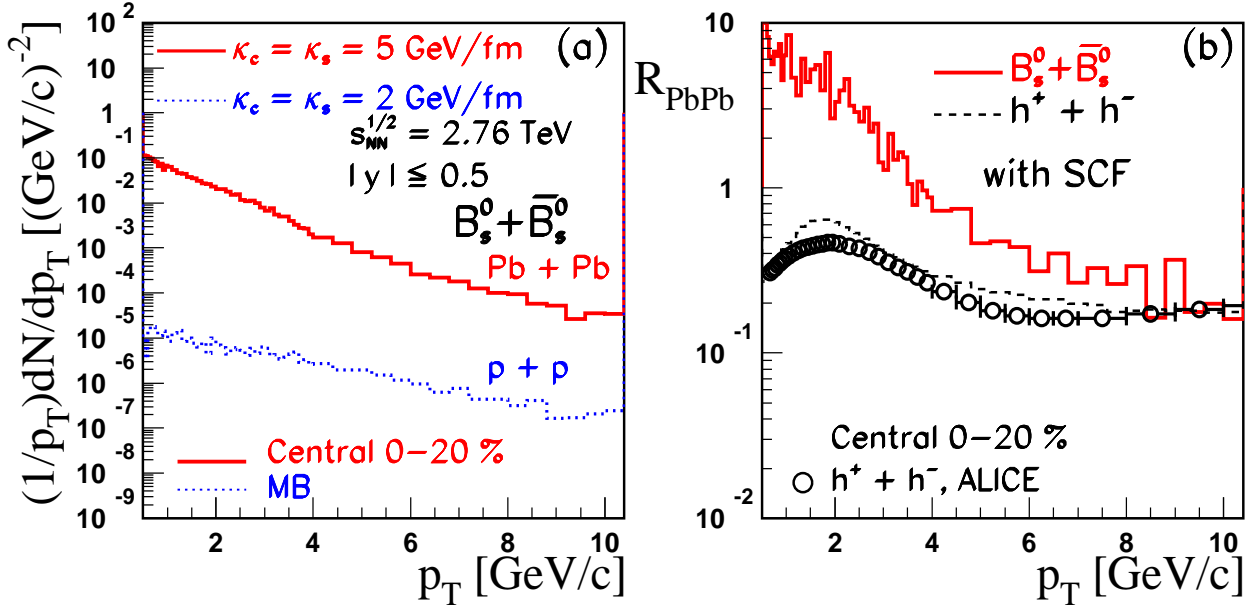}
\vskip 0.5cm
\caption[R_AA of Bs0 for Pb+Pb relative to pp] 
{\small (Color online) 
Part (a): {\small HIJING/B\=B v2.0} predictions for $p_T$ distributions at 
mid-rapidity for Pb + Pb $\rightarrow$ 
(B$^0_s$ + \=B$^0_s$)/2 + X (upper histogram), 
and for $p+p$ collisions (lower histogram).
Part (b): The $p_T$ dependence of NMF ${\rm R}_{\rm AA}({\rm p}_{\rm T})$ 
for B$^0_s$ mesons (solid histogram)  
and charged particles (dashed histogram) in central (0-20 \%) Pb + Pb collisions. Data are from ALICE Collaboration 
for charged particles (open circles) \cite{Abelev:2012hxa}.
Error bars include only statistical uncertainties.  
\label{fig:fig7}}
\end{figure}

In Fig.~\ref{fig:fig6} and  Fig.~\ref{fig:fig7} we address the beauty ($b$) 
quark production including results for non-strange $B^0$ and 
strange $B_s^0$ mesons. In the setup we kept the same parameters 
used for SCF ({\it i.e., $\kappa_b = \kappa_c= \kappa_s = 5 $} GeV/fm) and we have used the bottom mass M$_b^{\rm eff}$ = 4.16 GeV~\cite{Steinhauser:2008pm}.
The results of nuclear modification factor display a bump in the 
$p_T$ range 0.5-4 GeV/{\it c } with $R_{AA} > 1$ and a depletion at high $p_T$.
Since the quark mass play a negligible role at very large $p_T$, 
the model predicts the same supression for charm, bottom and and light quarks.
On the other hand, at small and moderate $p_T$, the bump mainly due to 
SCF effects is modified in amplitude and increase with increasing of the  
quark mass. Such a non-trivial behaviour at low $p_T$ if confirmed by experimental data, could be a crucial test for the role of SCF effects on heavy quark 
production at the LHC.  

The yields of strange mesons B$_s^0$ ( Fig.~\ref{fig:fig7}) are enhanced due to an increase  of $b\bar{b}$  and  $s\bar{s}$ pairs 
production (see Eq.~\ref{eq:gammaQ}).
In the moderate range of the transverse momentum the model predicts 
a quark-mass hierarchy, {\it i.e.}, 
$R_{\rm PbPb}^{\pi} \, < \,  R_{\rm PbPb}^{ch} \, < \,R_{\rm PbPb}^{B^0}<\,R_{\rm PbPb}^{B_s^0}$ , similar with those seen for charmed mesons. 

\subsection{D mesons ratios}

\begin{figure} [htb!]
\centering
\includegraphics[width=\linewidth,height=8.0cm]{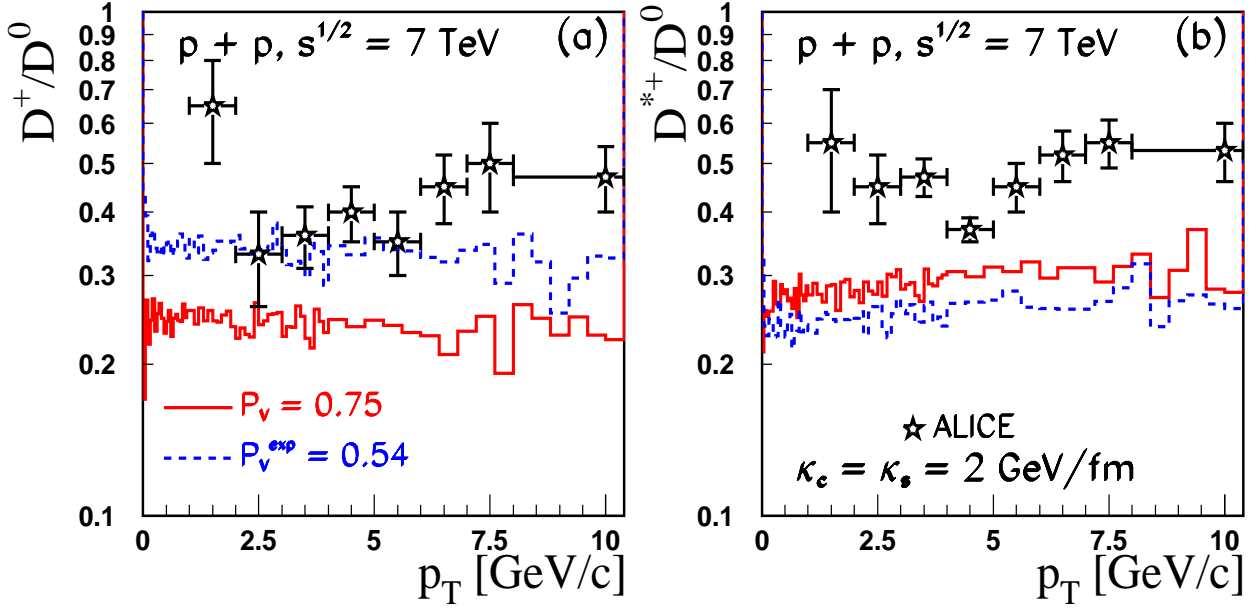}
\vskip 0.5cm\caption[ratio of non-strange mesons for p + p ] 
{\small (Color online) Comparison of {\small HIJING/B\=B v2.0} predictions 
for ratios of non-strange D mesons;
D$^+$/D$^0$ (left panel) and D$^{*+}$/D$^0$ (right panel). 
 Two sets of results are shown,
corresponding to default fraction P$_V$ = 0.75 solid histograms   
and for the measured fraction P$_V^{\rm exp}$ = 0.54  dashed histograms
(see text for explanation). The data are from Refs.~\cite{ALICE:2011aa}, \cite{Abelev:2012tca}.
Error bars include only the statistical uncertainty.
\label{fig:fig8}}
\end{figure}

The inclusive $p_T$ distributions for  
open prompt charmed mesons production (D$^0$, D$^+$, D$^{*+}$, D${_s}{^+}$) in $p + p$
collisions at $\sqrt{s}$ = 7 TeV were shown in Fig.~\ref{fig:fig1}. As 
noted in the caption of Fig.~\ref{fig:fig1} 
the reported yields refer to particles only, being computed as the average of
particles and antiparticles, in order to improve statistical uncertainties.
This assume that the production cross section is the same for 
particle (D) and antiparticle (\=D).
The {\small HIJING/B\=B} v2.0 model predictions for 
the $p_T$ dependence of ratios for non-strange mesons D$^+$ and D$^{*+}$
to that of  D$^0$ are shown in Fig.~\ref{fig:fig8}.
For comparison with data only 
D mesons in the rapidity range $|y| < 0.5$ were considered.

\begin{figure} [htb!]
\centering

\includegraphics[width=\linewidth,height=8.0cm]{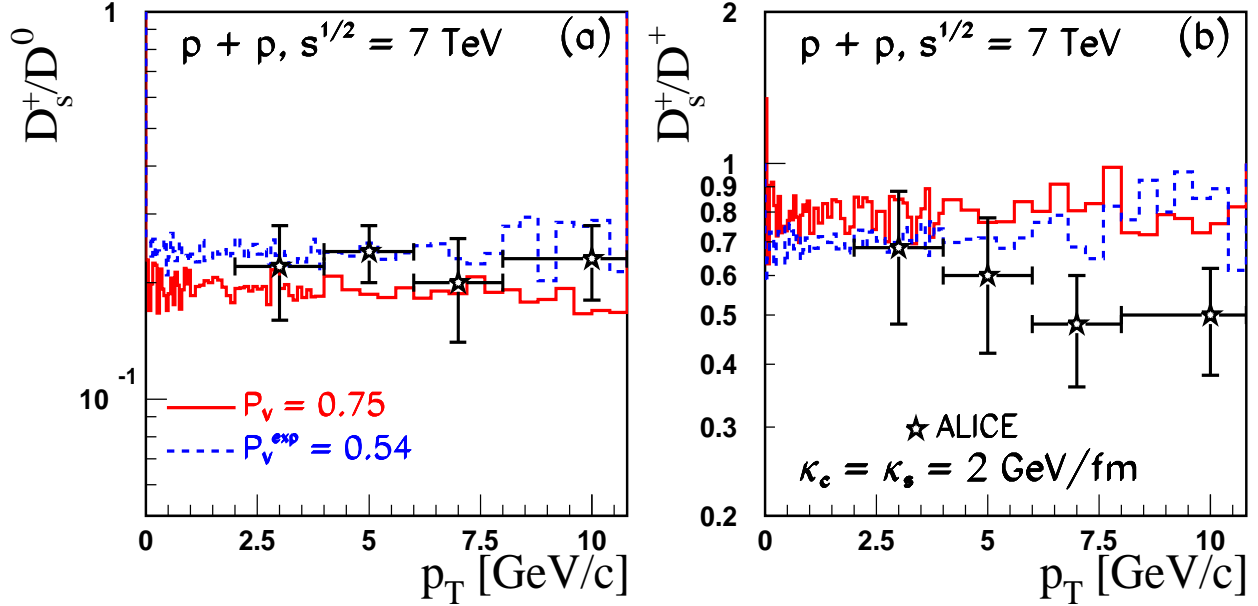}
\vskip 0.5cm\caption[ratio of non-strange mesons for p + p ] 
{\small (Color online) Comparison of {\small HIJING/B\=B v2.0} predictions 
for ratios of strange D${_s}{^+}$ to non-strange mesons D$^0$ (left panel) 
and D$^+$ (right panel). 
The histograms have the same meaning as in Fig.~\ref{fig:fig8}.
The data are from Refs.~\cite{ALICE:2011aa}, \cite{Abelev:2012tca}.
Error bars include only the statistical uncertainty.
\label{fig:fig9}}
\end{figure}

The D$^+$/D$^0$ and D$^{*+}$/D$^0$ ratios are determined in the model by an input parameter 
P$_V$ = $V/(V + S)$, that defines the fraction of 
D-mesons in vector state (V) to all produced mesons (vectors (V) + scalars (S)). 
The solid histograms in Fig.~\ref{fig:fig8} are obtained 
with the default value based on spin counting stattistics ({\it i.e.}, P$_V$ = 3/(3 + 1) = 0.75).
Taking rather for P$_V$ a value from the measured fractions
of heavy flavour mesons produced in a vector state  P$_V^{\rm exp.}$ = 0.54~\cite{Abelev:2012tca},
results in an enhancement of the  D$^+$/D$^0$ ratio (left panel) and a reduction of D$^{*+}$/D$^0$ ratio (right panel) 
as compared to those obtained with the P$_V$ default value.
The agreement with data is improved for the D$^+$/D$^0$ ratio. On the other hand,
the D$^{*+}$/D$^0$ ratio is underestimated by a factor of $\approx$ 1.5,
since the model predict a smaller cross sections for resonance  production of D$^{*+}$ meson in $p+p$ collisions at $\sqrt{s}$ = 7 TeV.
 
The ratios of prompt strange-meson  D${_s}{^+}$ to the  
non-strange meson D$^0$ and D$^+$ are plotted in Fig.~\ref{fig:fig9} .
These ratios are mainly controlled by another input parameter $\gamma_s$,
that defines the $s/u$ quark suppression factor in the fragmentation process.
In the {\small HIJING/B\=B} v2.0 model this parameter is set to 
 $\gamma_s$ = 0.45 using an energy dependent $\kappa$ in $p + p$ collisions,
and leads to an enhanced production of D${_s}{^+}$ mesons,
when compared with those using the default value  $\gamma_s$ = 0.3.
Note that $\gamma_s$ = 0.45 it is compatible within total uncertainties
with the measured values~\cite{Abelev:2012tca},  
$\gamma_s^{\rm exp}$ = $0.31 \pm 0.08$(stat)$\pm \, 0.10$(sys) $\pm \,0.02$(BR);
here BR stands for decay branching ratios.

\begin{figure} [h!]
\centering
\includegraphics[width=\linewidth,height=8.0cm]{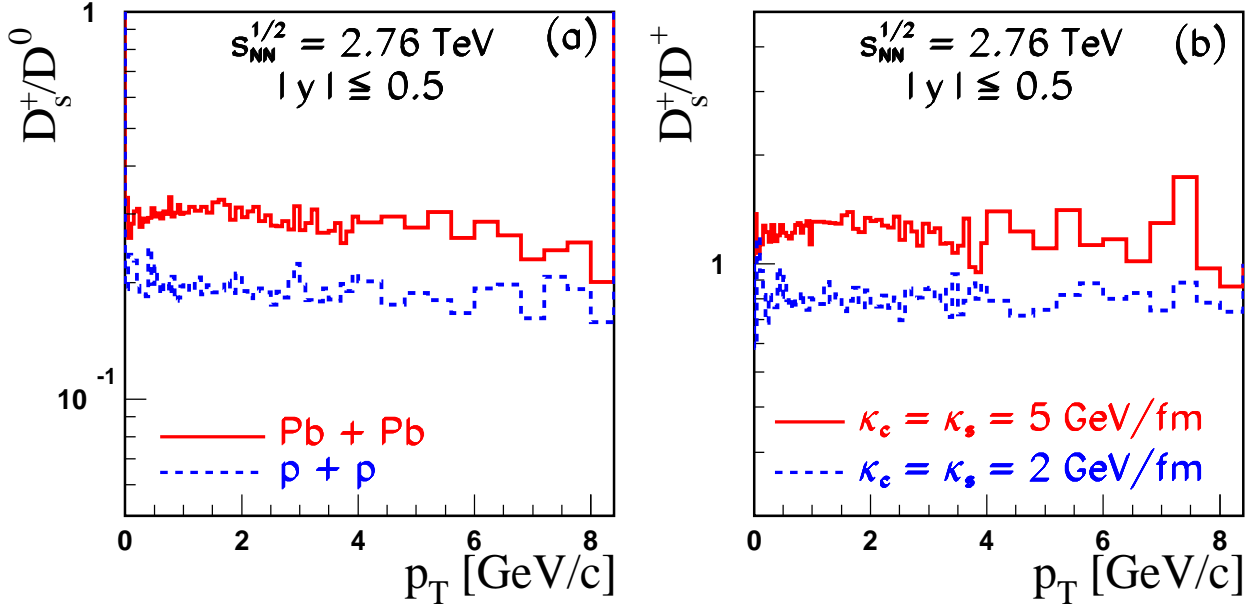}
\vskip 0.5cm\caption[ratio of non-strange mesons for p + p ] 
{\small (Color online) Comparison of {\small HIJING/B\=B v2.0} predictions 
for ratios of strange D${_s}{^+}$ to non-strange mesons D$^0$ (left panel) 
and D$^+$ (right panel) at $\sqrt{s}$ = 2.76 TeV in $p+p$ collisions  
and at $\sqrt{s_{\rm NN}}$ = 2.76 TeV
in centrality class 0-20 \% in Pb+Pb collisions.
The results are shown for $\kappa_c = \kappa_s = 2$ GeV/fm (dashed histogram) 
and for in medium value $\kappa_c = \kappa_s $= 5 GeV/fm (solid histograms).
The parameter P$_V$ is set at its default value P$_V$ = 0.75.
\label{fig:fig10}}
\end{figure}

The calculations describe fairly well the D${_s}{^+}$/D$^0$ ratio, while slightly overestimating the D${_s}{^+}$/D$^+$ ratio.
These ratios show almost no $p_T$ dependence 
due to a very small difference between the fragmentation function 
of charm quarks to strange and non-strange mesons. 
Note that PYTHIA with Perugia-0 tune (using  $\gamma_s$ = 0.2) underestimates the 
strange prompt meson production~\cite{Abelev:2012tca}.
More precise data are clearly needed to reach a firmer conclusion.

It will be interesting to study whether the ratios 
of strange to non-strange charmed mesons {\it i.e}, 
D${_s}{^+}$/D$^0$ and  D${_s}{^+}$/D$^+$ 
are enhanced in central Pb + Pb collisions
relative to $p+p$ collisions. In Fig.~\ref{fig:fig10} are shown the calculated 
ratios obtained at the same centre of mass  energy. 
The calculations are performed for $\kappa_c = \kappa_s = 2$ GeV/fm 
in $p+p$ collisions (dashed histograms) 
and for in medium value $\kappa_c = \kappa_s $= 5 GeV/fm (solid histograms).  
An enhancement of a factor of $\approx$ 2  is predicted
by the {\small HIJING/B\=B} v2.0 model 
in going from $p+p$ minimum bias events to central Pb + Pb collisions.
If the data confirm this enhancement, then one could conclude that
the assumption of in medium increase of {\em the effective} string tension or 
equivalently ``{\em in-medium mass modification}'' of charm quark, due to 
possible induced chiral symmetry restoration, is supported.

\section{Summary and Conclusions}

In summary, we studied the influence of possible strong homogeneous 
constant color electric fields on open prompt charmed mesons (D$^0$, D$^+$, D$^{*+}$, 
D${_s}{^+}$) production in Pb + Pb and minimum bias events $p + p $ collisions
in the framework of the {\small HIJING/B\=B v2.0} model. 
The measured ratios of prompt strange-meson  D${_s}{^+}$ to the  
non-strange meson D$^0$ and D$^+$ in minimum bias $p+p$ collisions at $\sqrt{s}$ = 7 TeV
help to verify our assumptions 
and to set the strangeness suppression factor for charm mesons. 
We assume an energy and system dependence of the effective string tension,
$\kappa$, equivalent with an {\em in medium mass} modification 
of charm and strange quark. 
The effective string tension control Q\=Q pair creation rates 
and suppression factors $\gamma_{Q\bar{Q}}$.   

For Pb + Pb collisions at $\sqrt{s_{\rm NN}}$ = 2.76 TeV  
all nuclear effects included in the model, {\it e.g.}, 
strong color fields, shadowing and quenching should be taken into account.
Partonic energy loss and jet quenching process as embedded in the model 
achieve a reasonable description of the suppression ($R_{\rm PbPb}^{\rm D} < 1 $)
at moderate and high transverse momentum.
Moreover, at low and intermediate $p_T$ ($0 < p_T < 8 $ GeV/{\it c}) 
the model predicts a quark mass hierarchy as suggested in Ref.~\cite{Armesto:2005iq}.
By computing nuclear modification factor $R_{\rm PbPb}^{\rm D}$, 
we show that the above nuclear effects constitute important 
dynamical mechanisms that explain better the observed 
prompt {\it D}-mesons and charged particles production as observed by the ALICE collaboration.

The initial production of $c\,\bar{c}$ pairs by gluon fusion might be suppressed due to initial state effects ({\it e.g.} gluon shadowing or saturation).
By computing the nuclear modification factor $R_{\rm pPb}^{\rm D}$ in 
central $p$ + Pb collisions at $\sqrt{s_{\rm NN}}$ = 5.02 TeV 
including shadowing effects, we conclude that the strong 
suppression observed for $R_{\rm PbPb}^{\rm D}$ is due to a final state effect.
Measurements with high statistics at low $p_T$ 
 ($0 < p_T < 4 $GeV/{\it c}) of the nuclear modification factor 
$R_{\rm PbPb}^{\rm D}$ and $R_{\rm PbPb}^{\rm B}$ in Pb + Pb central collisions, could help to disentangle  
between different model approaches and/or different dynamical mechanisms,
especially for D$_s^+$ ($c\bar{s}$) and B$^0_s$ ($b\bar{s}$) mesons, due to their quark 
content.

The {\small HIJING/B\=B model} is based on a time-independent 
strength of color field, while in 
reality the production of Q\=Q pairs is more complex being far-from-equilibrium,
time and space dependent phenomenon. To achieve more quantitative
conclusions, such time and space dependent mechanisms 
\cite{Hebenstreit:2008ae,Levai:2009mn}
should be considered in future generations of Monte Carlo codes.

\section{Acknowledgments}
\vskip 0.2cm

This work is supported by the Natural Sciences and Engineering 
Research Council of Canada (V.~Topor~Pop, J.~Barrette, and C.~Gale),
by the Division of Nuclear Science, 
U.S. Department of Energy, under Contract No. DE-AC03-76SF00098 and
DE-FG02-93ER-40764 (associated with the JET Topical Collaboration Project, M.~Gyulassy), and 
by the Romanian Authority 
for Scientific Research, CNCS-UEFIS-CDI project 
number PN-II-ID-2011-3-0368 (M.~Petrovici, and ~V.~Topor~Pop partial support).

\end{document}